\documentclass[modern,trackchanges]{aastex63}

\usepackage{microtype}
\usepackage{url}
\usepackage{amsmath}
\usepackage{mathtools}
\usepackage{esint}
\usepackage{amssymb}
\usepackage{natbib}
\usepackage{multirow}
\usepackage{graphicx}
\usepackage{scalerel}
\usepackage{calc}
\usepackage{etoolbox}
\usepackage{marginnote}
\usepackage{nicefrac}
\usepackage{tabstackengine}
\usepackage{diagbox}
\usepackage[makeroom]{cancel}
\usepackage{mathdots}
\usepackage{bbm}
\usepackage{booktabs}
\usepackage{xspace}
\usepackage{upgreek}
\usepackage[T1]{fontenc} 
\usepackage{textcomp}
\usepackage{ifxetex}
\ifxetex
\usepackage{fontspec}
\defaultfontfeatures{Extension = .otf}
\fi
\usepackage{fontawesome}
\usepackage{listings}
\usepackage{mathtools}
\stackMath

\newcommand{\starry}{\texttt{starry}\xspace}
\newcommand{\exoplanet}{\texttt{exoplanet}\xspace}
\newcommand{\batman}{\texttt{batman}\xspace}

\newcommand{\Python}{\texttt{Python}\xspace}

\newcommand{\thiswork}{Agol, Luger, \& Foreman-Mackey (2019)\xspace}

\renewcommand{\eqref}[1]{\ref{eq:#1}}
\newcommand{\Eq}[1]{Equation~(\eqref{#1})}
\newcommand{\eq}[1]{\Eq{#1}}

\definecolor{linkcolor}{rgb}{0.1216,0.4667,0.7059}
\newcommand{\codeicon}{{\color{linkcolor}\faFileCodeO}}
\newcommand{\prooficon}{{\color{linkcolor}\faPencilSquareO}}
\newcommand{\animicon}{{\color{linkcolor}\faPlayCircle}}
\newcommand{\pycodelink}[1]{\href{https://github.com/rodluger/Limbdark.jl/blob/arxiv-v2/tex/figures/python/#1.py}{\codeicon}\,\,}
\newcommand{\jlcodelink}[1]{\href{https://github.com/rodluger/Limbdark.jl/blob/arxiv-v2/tex/figures/julia/#1.jl}{\codeicon}\,\,}
\newcommand{\animlink}[1]{\href{https://github.com/rodluger/Limbdark.jl/blob/arxiv-v2/tex/figures/#1.gif}{\animicon}\,\,}
\newcommand{\prooflink}[1]{\href{https://github.com/rodluger/Limbdark.jl/blob/arxiv-v2/proofs/#1.ipynb}{\raisebox{-0.1em}{\prooficon}}}

\newtagform{eqtag}[]{(}{)}
\newcommand{\currentlabel}{None}

\newenvironment{proof}[1]{%
\ifstrempty{#1}{%
\renewtagform{eqtag}[]{\raisebox{-0.1em}{{\color{red}\faPencilSquareO}}\,(}{)}%
}{%
\renewtagform{eqtag}[]{\prooflink{#1}\,(}{)}%
}%
\usetagform{eqtag}%
\renewcommand{\currentlabel}{#1}
\align%
}{%
\endalign%
\renewtagform{eqtag}[]{(}{)}%
\usetagform{eqtag}%
\message{<<<\currentlabel: \theequation>>>}%
}

\newenvironment{proof*}[1]{%
\ifstrempty{#1}{%
\renewtagform{eqtag}[]{\raisebox{-0.1em}{{\color{red}\faPencilSquareO}}\,(}{)}%
}{%
\renewtagform{eqtag}[]{\prooflink{#1}\,(}{)}%
}%
\usetagform{eqtag}%
\renewcommand{\currentlabel}{#1}
\equation%
}{%
\endequation%
\renewtagform{eqtag}[]{(}{)}%
\usetagform{eqtag}%
\message{<<<\currentlabel: \theequation>>>}
}

\newcommand{\T}{\ensuremath{\mathrm{T}}}
\newcommand{\dd}{\ensuremath{ \mathrm{d}}}

\newcommand{\bvec}[1]{{\ensuremath{\mathbf{#1}}}}

\newcommand{\x}{\ensuremath{\mbox{$x$}}}
\newcommand{\y}{\ensuremath{\mbox{$y$}}}
\newcommand{\z}{\ensuremath{\mbox{$z$}}}
\newcommand{\xhat}{\ensuremath{\mathbf{\hat{x}}}}
\newcommand{\yhat}{\ensuremath{\mathbf{\hat{y}}}}

\DeclareMathAlphabet\mathbfcal{OMS}{cmsy}{b}{n}

\definecolor{dim}{rgb}{0.8,0.8,0.8}
\newcolumntype{L}[1]{>{\raggedright\let\newline\\\arraybackslash\hspace{0pt}}m{#1}}
\newcommand{\bigdot}{\scaleto{\cdot}{6pt}}

\newcommand{\ubasis}{\ensuremath{\tilde{\mathbf{u}}}}

\newcommand{\pbasis}{\ensuremath{\tilde{\mathfrak{p}}}}

\newcommand{\gbasis}{\ensuremath{\tilde{\mathfrak{g}}}}
\newcommand{\gbasisn}{\ensuremath{\tilde{\mathfrak{g}}_n}}

\usepackage{listings}
\definecolor{codegreen}{rgb}{0,0.6,0}
\definecolor{codegray}{rgb}{0.5,0.5,0.5}
\definecolor{codepurple}{rgb}{0.58,0,0.82}
\definecolor{backcolour}{rgb}{0.95,0.95,0.95}
\lstdefinestyle{mystyle}{
    backgroundcolor=\color{backcolour},
    commentstyle=\color{codegreen},
    keywordstyle=\color{magenta},
    numberstyle=\tiny\color{codegray},
    stringstyle=\color{codepurple},
    basicstyle=\small\ttfamily,
    breakatwhitespace=false,
    breaklines=true,
    captionpos=b,
    keepspaces=true,
    numbers=left,
    numbersep=5pt,
    showspaces=false,
    showstringspaces=false,
    showtabs=false,
    tabsize=2,
    aboveskip=1em,
    belowskip=1em,
    keywords=[2]{map},
    keywordstyle=[2]{\color{black!80!black}},
}
\makeatletter
\def\Ddots{\mathinner{\mkern1mu\raise\p@
\vbox{\kern7\p@\hbox{.}}\mkern2mu
\raise4\p@\hbox{.}\mkern2mu\raise7\p@\hbox{.}\mkern1mu}}
\makeatother

\renewcommand\quad{\hskip\fontdimen3\font}

\newcommand{\edited}{}

\usepackage{graphicx}

\begin{document}

\setlength{\abovedisplayskip}{1.5em}
\setlength{\belowdisplayskip}{1.5em}

\title{%
Analytic Planetary Transit Light Curves and Derivatives for Stars with Polynomial Limb Darkening
}

\author[0000-0002-0802-9145]{Eric Agol}\altaffiliation{Guggenheim Fellow}
\affil{Department~of~Astronomy, University~of~Washington, Seattle, WA}
\affil{Virtual~Planetary~Laboratory, University~of~Washington, Seattle, WA}
\author[0000-0002-0296-3826]{Rodrigo Luger}
\affil{Center~for~Computational~Astrophysics, Flatiron~Institute, New~York, NY}
\affil{Virtual~Planetary~Laboratory, University~of~Washington, Seattle, WA}
\author[0000-0002-9328-5652]{Daniel Foreman-Mackey}
\affil{Center~for~Computational~Astrophysics, Flatiron~Institute, New~York, NY}


\begin{abstract}
We derive analytic, closed-form solutions for the light curve of a planet
transiting a star with a limb darkening profile which is a polynomial function
of the stellar elevation, up to arbitrary integer order.
We provide improved analytic expressions for the uniform, linear, and quadratic
limb-darkened cases, as well as novel expressions for higher order integer powers
of limb darkening.  The formulae are crafted to be numerically stable over the
expected range of usage.  We additionally present analytic formulae for
the partial derivatives of instantaneous flux with respect to the radius ratio,
impact parameter, and limb darkening coefficients.  These expressions are rapid to
evaluate, and compare quite favorably in speed and accuracy to existing transit light
curve codes. We also use these expressions to numerically compute the first partial
derivatives of exposure-time averaged transit light curves with respect to all
model parameters.  An additional application is modeling eclipsing binary or
eclipsing multiple star systems in cases where the stars may be treated as spherically
symmetric.  We provide code which implements these formulae in \texttt{C++}, 
\texttt{Python}, \texttt{IDL}, and \texttt{Julia},
with tests and examples of usage.
 \href{https://github.com/rodluger/Limbdark.jl}{\color{linkcolor}\faGithub}
\end{abstract}

%
\section{Introduction}
\label{sec:intro}

The precise measurement of the transits of an exoplanet offers a host of information
about the planet's properties \citep{Charbonneau2007,Winn2008,Winn2010,Haswell2010}.  To start with, 
the times of transit give the planet's orbital ephemeris.  The depth of transit, corrected 
for stellar limb darkening, gives the planet's radius relative to that of the star
{\edited \citep{Heller2019}}.   The
shape of the transit, especially the duration of ingress and egress relative to
the full transit duration, yields the orbital impact parameter of the planet, which
constrains the inclination of the orbit relative to the observer \citep{Seager2003}.
Beyond these basic properties, if the transit depth is seen to vary with wavelength,
the presence of spectral features may be used to constrain the chemical composition
of the planet's atmosphere \citep{Brown2001,Seager2010,Burrows2014,Crossfield2015,
Madhusudhan2019}.  If the transit times are seen to vary, a dynamical
model can constrain the masses of the planet companions, and vice versa \citep{Agol2005,Holman2005}.  If
the planet is seen in eclipse, its temperature, emission spectrum, and atmospheric
circulation pattern can be constrained \citep{Cowan2017,Alonso2018}.  When combined with radial velocity
measurements, the bulk density of a planet can be inferred, yielding constraints
on its bulk composition \citep{Udry2007}.

And yet, all of these inferences are predicated on the precise computation of models
of the planetary transit which may be used to infer the model parameters.  Stars
are non-uniform in brightness, with the general
trend of growing dimmer towards the limb, and so limb darkening must be accounted
for to accurately infer the planetary parameters \citep{Csizmadia2018}.  Indeed, fast
and accurate computation of limb-darkened transit light curve models has enabled the
detection and characterization of thousands of transiting exoplanets \citep{MandelAgol2002}.
The most important ingredient to these models has been a description of the
limb darkening model which is flexible enough and accurate enough to describe the
emission from a stellar photosphere.  Linear and quadratic limb darkening laws
were sufficient for lower-precision measurements; however, the measurement of
transit light curves has steadily improved in precision.  Higher order terms or non-linear laws
have become necessary to describe higher precision measurements \citep{Kopal1950,Claret2000,
Gimenez2006}, which tend to involve more computational burden.

In addition to computing transit light curves, the derivatives of these light
curves with respect to the model parameters are also beneficial for accurate
characterization of exoplanets.   The derivatives enable fast and stable
optimization of the transit light curve parameters, which is critical for
obtaining initial estimates for a Markov Chain Monte Carlo simulation
\citep[MCMC; e.g.][]{Ford2005,Ford2006},
for looking for multi-modal solutions, for initializing the multi-nest
algorithm \citep{Feroz2008}, or for computing the Fisher information matrix 
\citep{Vallisneri2008}.
In some cases, MCMC can be slow to converge, and derivatives can accelerate
convergence by adding an artificial momentum term to the log likelihood,
and then allowing the sampler to follow contours of constant ``energy.''
This so-called ``hybrid'' or ``Hamiltonian'' MCMC approach holds great promise \citep{Neal2011,Girolami2011,Betancourt2017},
but its application has been hampered by the lack of models with derivatives,
as derivatives are in general more difficult to compute.

Finally, the analytic\footnote{By ``analytic" we mean closed-form, not
infinitely differentiable.} computation of transit light curves with quadratic
limb darkening has a precision which can be limited by numerical round-off
error for parameters near some special cases.  In
particular, when the radius equals the impact-parameter, which corresponds to
the edge of the planet crossing the center of the star, the computation of the
elliptic integrals becomes unstable.  At the second and third points of
contact, when the radius of the planet plus the impact parameter equals
the radius of the star, the elliptic integrals diverge logarithmically.  In the
limit that the impact parameter approaches zero, the equations can also
diverge.  All of these special cases are in principle encountered rarely,
but in practice with thousand of planets with tens to thousands of
transits each, along with hundreds to hundreds of thousands of light
curves with time sub-sampled for each exposure, these rare cases can
be encountered with some frequency.

Based on these considerations, the primary goals of the current paper are
threefold:
\begin{enumerate}
\item To extend the analytic quadratic transit model to higher order limb darkening.
\item To compute the derivatives of the model analytically.
\item To stabilize the analytic light curve computation (and its derivatives)
in all limits near special cases.
\end{enumerate}
Secondary goals include modeling eclipsing binaries, for which the same
considerations apply, and integrating the light curve model, and its
derivatives, quickly and accurately over time to account for finite
exposure times.

Some progress has been made already towards these goals.  To describe this
progress, we pause first to introduce some notation.  Limb darkening models
of spherical stars are parameterized with the cosine of the angle measured
from the sub-stellar point, $\upmu = \cos{\theta}$, where $\theta$ is the
polar angle on the photosphere, with $\theta=0$ at the center of the observed
stellar disk, and $\theta=\pi/2$ at the limb.  In a coordinate system in which
the projected disk of the star lies in the $x-y$ plane, and the $z$ coordinate
points towards the observer, then $\upmu = z$, where $x$, $y$, and $z$ are 
measured in units of the stellar radius.  The variable $\upmu=z$ is then the 
elevation on the surface of the star, where the highest point is taken to be 
closest to the observer.  In terms of $b=\sqrt{x^2+y^2}$, the normalized separation projected onto
the sky, this parameter is given by $\upmu =\sqrt{1-b^2}$, where
$0\le b \le 1$ within the stellar disk.  We also introduce the radius ratio,
$r$, which is the radius of the occultor divided by the radius of the source.
In general, we will follow the notation
introduced by \citet{starry} for the \starry code package.

Uniform limb darkening scales as $I(\upmu) \propto \upmu^0$, first-order
limb darkening as $I(\upmu) \propto \upmu^1$, and second-order limb darkening
as $I(\upmu)\propto \upmu^2$; these are the three most commonly used  terms
that can be integrated analytically, which we describe in detail below
in sections \ref{sec:uniform}, \ref{sec:reparam}, and \ref{sec:quadratic}.
These are typically combined to yield the quadratic limb darkening law,
\begin{equation} \label{eq:quadraticld}
    \frac{I(\upmu)}{I_0} = 1-u_1 (1-\upmu) - u_2 (1-\upmu)^2,
\end{equation}
where $u_1$ and $u_2$ are the limb darkening parameters, and
$I_0 \equiv I(1)$ is a normalization constant, equal to the
intensity at the center of the stellar disk.
In this paper, we will show that higher order powers of $\upmu^n$ with integer
$n$ can be integrated analytically for $n > 2$ when expressed as recursion relations.
Linear combinations of these laws can be constructed,
with various parameterizations, to describe stellar limb darkening more precisely.

The first goal of modelling higher-order limb darkening was accomplished
by \citet{Gimenez2006}, who derived transit light curves for a limb darkening
function
\begin{equation} \label{eq:gimenez}
    \frac{I(\upmu)}{I_0} = 1-\sum_{n=1}^N a_n (1-\upmu^n) \quad,
\end{equation}
where $a_n$ is a limb darkening coefficient.  \cite{Gimenez2006}
found an infinite series expansion for computing the limb-darkened light curve
for each $a_n$ term.  {\edited This algorithm is remarkable in that it allows
for computation of limb-darkening to arbitrary polynomial order, and gives
excellent single-precision accuracy and better speed than numerical integration
approaches.}  Here we {\edited improve upon the pioneering work of \citet{Gimenez2006}
by presenting} closed-form expressions for these terms
which can be easily computed with recursion relations, although for purposes
of numerical stability we need to revert to series solutions in some limits
which we find to be rapid to evaluate.  {\edited In addition to being faster
to evaluate and more accurate for low-order limb-darkening 
(\S \ref{sec:comparison_pytransit}), these new expressions
also include derivatives with respect to the model parameters.}

The second goal, of computing derivatives of the light curve with respect to
the model parameters, was accomplished by \cite{Pal2008} for the quadratic
limb darkening case.  P\'al derived the partial derivatives of the quadratic
limb darkening model with respect to $b$, $r$, and the two quadratic
limb darkening coefficients. In this work, we give modified expressions
for the quadratic limb-darkened flux and its derivatives which are
more numerically stable, as well as extend the computation of derivatives
to higher order limb darkening.

The third goal, of numerical stability, has yet to be addressed in the literature.
Although some numerical approaches are numerically stable, such as \cite{Gimenez2006},
\cite{Kreidberg2015}, and \cite{Parviainen2015}, these approaches tend to be
slower, they have precisions which may depend upon the tolerance of the computation
which is specified, and, in addition, they do not yield derivatives of the light curves.
The expressions presented in this work were derived with numerical stability in
mind, and we show that for low-order limb darkening our expressions are accurate
to {\edited double} precision in nearly all cases.

A disadvantage of our approach is that it requires integer powers of the limb darkening
expansion.  \citet{Claret2000} has shown that a non-linear limb darkening law,
with half-integer powers of $\upmu$, gives an accurate description of stellar
limb darkening models.  More recently, the power-law model, $I(\upmu) = 1-
c_\alpha(1-\upmu^\alpha)$ \citep{Hestroffer1997} was shown to be an accurate
limb darkening law despite only using two parameters \citep{Morello2017,Maxted2018}.
We were unable to find an analytic solution for these limb darkening
laws, but we will compare with these models below in \S \ref{sec:comparison}.



This paper is organized as follows. In \S\ref{sec:poly_limbdark}
we introduce the general form for polynomial limb darkening
and define the notation used throughout the paper. In \S\ref{sec:uniform}--\ref{sec:quadratic}
we derive updated equations for the well-known cases of uniform, linear, and
quadratic limb darkening, and in \S\ref{sec:higher_order} we generalize the
expressions to limb darkening of arbitrary order.
We discuss time integration of the equations (for finite exposure time)
in \S\ref{sec:time}, an application to modeling non-linear limb
darkening in \S\ref{sec:nonlinear}, and details on the implementation
of the algorithm in \S\ref{sec:implementation}. In \S\ref{sec:benchmark} and
\S\ref{sec:comparison} we discuss timing benchmarks and comparisons to
existing codes. Finally, in \S\ref{sec:discussion}--\S\ref{sec:conclusions}
we discuss our assumptions, caveats of our modeling, applications of our
algorithm, and a summary of our results. Appendices A--C contain a list
of errata for \citet{MandelAgol2002}, derivatives of the general
complete elliptic integral, and a comprehensive list of symbols used in
the paper.

Finally, as in \cite{starry} and \cite{AprilFools}, we embed links
to \Python code (\,\codeicon\,) to reproduce all of the
figures, as well as links to \texttt{Jupyter} notebooks
(\,\prooficon\,) containing proofs and derivations
of the principal equations. We urge members of the community to do the same
to improve the accessibility, transparency, and reproducibility of
research in astronomy.

%

\section{Polynomial Limb Darkening}
\label{sec:poly_limbdark}

In analogy with the quadratic
limb darkening law (Equation~\ref{eq:quadraticld}), let us define the
generalized polynomial limb darkening law of order $N$ as
\begin{align}
    \label{eq:polynomialld}
    \frac{I(\upmu)}{I_0} &= 1 - u_1 (1 - \upmu) - u_2 (1 - \upmu)^2 -
                                ... - u_{N}(1 - \upmu)^{N} \nonumber \\
                          &= -\sum_{i=0}^N u_i (1 - \upmu)^i
\end{align}
%
where we define $u_0 \equiv -1$. In a right-handed Cartesian coordinate system centered
on the body, with the $z$-axis pointing to the observer,
\begin{equation}\label{eq:xyz}
\upmu(x, y) = z(x, y) = \sqrt{1 - x^2 - y^2}.
\end{equation}
%
%
If we let $\bvec{u}$ be the column vector of limb darkening coefficients
$\bvec{u} \equiv (u_0 \ u_1 \ u_2 \ ... \ u_N)^\mathsf{T}$
and $\ubasis$ be the \emph{limb darkening basis}
\begin{align}
    \label{eq:ldbasis}
    \ubasis = -\begin{pmatrix}
        1 & &
        (1 - z) & &
        (1 - z)^2 & &
        ... & &
        (1 - z)^N
    \end{pmatrix}^\mathsf{T} \quad,
\end{align}
we may
express Equation~(\ref{eq:polynomialld}) more compactly as the
dot product
\begin{align}
    \label{eq:polynomialld_vec}
    \frac{I(z)}{I_0} &= \ubasis^\mathsf{T} \bvec{u} \quad .
\end{align}

In this paper, our task is to compute the flux, $F$, observed during a transit or occultation by
integrating this function over the visible area of the disk:
\begin{align}
    \label{eq:occint}
    F &=
    \iint I(z) \, \dd S \quad .
\end{align}
In general, the surface integral in Equation~(\ref{eq:occint}) is difficult---if not
impossible---to solve directly with $I(z)$ given by Equation~(\ref{eq:polynomialld_vec}).
However, as in \citet{starry}, we note that the problem
is made significantly more tractable if we first perform two change of basis
operations.

\subsection{Change of basis}
\label{sec:change_of_basis}
We wish to find a basis in which to express the limb darkening profile that
makes evaluating Equation~(\ref{eq:occint}) easier. This section follows
closely the discussion in \citet{starry}, in which the authors first transform
to a \emph{polynomial basis}, whose terms are simple powers of the coordinates,
and then to a \emph{Green's basis}, whose terms make application of Green's
theorem convenient in reducing the surface integral to a one-dimensional line
integral.

Let us define the transformation to the polynomial basis by the linear equation
\begin{align}
    \label{eq:pbasis}
    \mathfrak{p} = \mathcal{A}_1 \bvec{u}
\end{align}
where $\mathfrak{p}$ is the vector of limb darkening coefficients in the
polynomial basis $\pbasis$ and $\mathcal{A}_1$ is a change of basis matrix.
We define the polynomial basis to be the power series in $z$,
\begin{align}
    \label{eq:polybasis}
    \pbasis = \begin{pmatrix}
        1 & z & z^2 & z^3 & ... & z^N
    \end{pmatrix}^\mathsf{T} \quad,
\end{align}
so that $\pbasis^T \cdot \mathfrak{p} = \ubasis^T \cdot \bvec{u}$.


The transformation between vectors in $\ubasis$ and vectors in
$\pbasis$ is straightforward. By the binomial theorem, we may write
the $i^\mathrm{th}$ coefficient of $\mathfrak{p}$ as
\begin{equation}
    \label{eq:an_of_un}
    \mathfrak{p}_i = (-1)^{i+1}\sum_{j=0}^N \binom{j}{i} u_j.
\end{equation}
The elements of the matrix $\mathcal{A}_1$ are thus given by
\begin{align}
    \label{eq:A1}
    \mathcal{A}_{1_{i, j}} = (-1)^{i+1}\binom{j}{i} \quad .
\end{align}
Note that, as with the limb darkening basis,
the specific intensity at a point may be written
\begin{align}
    \label{eq:pbasis_intensity}
    \frac{I(z)}{I_0} &= \pbasis^\mathsf{T} \mathfrak{p} \nonumber \\
                     &= \pbasis^\mathsf{T} \mathcal{A}_1 \mathbf{u}  \quad .
\end{align}
Next, we transform to the Green's basis via the equation
\begin{align}
    \label{eq:gbasis}
    \mathfrak{g} = \mathcal{A}_2 \mathfrak{p}
\end{align}
where $\mathfrak{g}$ is the vector of limb darkening coefficients in the
Green's basis $\gbasis$ and $\mathcal{A}_2$ is another change of basis matrix.
For reasons that will become clear later in this paper,
we define the Green's basis to be
\begin{align}
    \gbasis_n &=
    \begin{dcases}
        1 & \qquad n = 0
        \\
        z & \qquad n = 1
        \\
        (n+2)z^n-n z^{n-2} & \qquad n \ge 2
    \end{dcases}
    \nonumber\\[0.5em]
    \gbasis &=
    \begin{pmatrix}
        1 & &
        z & &
        4z^2 - 2 &&
        5z^3 - 3z &&
        ...
    \end{pmatrix}^\mathsf{T}
    \quad,
    \label{eq:greensbasis}
\end{align}
Given this definition, the columns of the change of basis matrix
$\mathcal{A}_2$ are the Green's basis vectors corresponding to each
of the polynomial terms in Equation~(\ref{eq:polybasis}). Note that
in practice, it is more efficient to transform vectors in the $\pbasis$
basis to vectors in the $\gbasis$ basis via the
downward recursion relation
\begin{equation}
    \label{eq:dn_of_an}
    \mathfrak{g}_n = 
\begin{dcases} 
\frac{\mathfrak{p}_n}{n+2} + \mathfrak{g}_{n+2} & \qquad N \ge n \ge 2\\
\mathfrak{p}_n + (n+2) \mathfrak{g}_{n+2} & \qquad n=1,0
\end{dcases},
\end{equation}
starting with $n=N$ and $\mathfrak{g}_{N+1}=\mathfrak{g}_{N+2}=0$.

As before, the specific intensity at a point may be written
\begin{align}
    \label{eq:pbasis_intensity}
    \frac{I(z)}{I_0} &= \gbasis^\mathsf{T} \mathfrak{g} \nonumber \\[0.5em]
                     &= \gbasis^\mathsf{T} \mathcal{A}_2 \mathfrak{p} \nonumber \\[0.5em]
                     &= \gbasis^\mathsf{T} \mathcal{A} \ \mathbf{u} \quad ,
\end{align}
where we define the complete change of basis matrix from limb darkening
coefficients to Green's coefficients
\begin{proof}{A}
    \label{eq:A}
    \mathcal{A} \equiv \mathcal{A}_2 \mathcal{A}_1 \quad.
\end{proof}
As an example, the full change of basis matrix for $N = 5$ is
\begin{proof}{A}
    \label{eq:Aexample}
    \mathcal{A} = \left(\begin{matrix}-1 & -1 & - \frac{3}{2} & - \frac{5}{2} 
    & - \frac{13}{3} & - \frac{23}{3}\\0 & 1 & 2 & \frac{18}{5} & \frac{32}{5} 
& \frac{80}{7}\\0 & 0 & - \frac{1}{4} & - \frac{3}{4} & - \frac{5}{3} 
& - \frac{10}{3}\\0 & 0 & 0 & \frac{1}{5} & \frac{4}{5} & \frac{15}{7}\\0 & 0 & 0 & 0 
& - \frac{1}{6} & - \frac{5}{6}\\0 & 0 & 0 & 0 & 0 & \frac{1}{7}\end{matrix}\right) \quad.
\end{proof}
The link next to Equation~(\ref{eq:Aexample}) provides code
to compute $\mathcal{A}$ for any value of $N$.
Finally, for future reference, for the common case of quadratic limb darkening, the
Green's vector is given by the dot product of $\mathcal{A}$ and the
vector of limb darkening coefficients and is equal to
\begin{proof}{A}
    \mathfrak{g} &= \mathcal{A} \bvec{u} \nonumber \\
                 &= \left(1-u_1-\tfrac{3}{2}u_2 \,\,\,\,\,\,\,\,\, u_1+2u_2 \,\,\,\,\,\,\,\,\, -\tfrac{1}{4}u_2\right)^\top \quad.
\end{proof}

\subsection{Computing the surface integral}
\label{sec:theintegral}

Given our reparametrization in terms of Green's polynomials, we may
re-write Equation~(\ref{eq:occint}) as
\begin{align}
    \label{eq:occint_greens}
    F &= \iint I(z) \, \dd S \nonumber \\[0.5em]
      &= I_0 \iint \gbasis^\mathsf{T} \mathcal{A} \ \bvec{u} \, \dd S \,  \nonumber \\[0.5em]
      &= I_0 \left( \iint \gbasis (z) \, \dd S \right)^\mathsf{T} \mathcal{A} \ \bvec{u} \nonumber \\[0.5em]
      &= I_0 \, \mathfrak{s}^\mathsf{T} \mathcal{A} \bvec{u} \quad ,
\end{align}
where
\begin{align}
    \label{eq:solution_vector}
    \mathfrak{s} \equiv \iint \gbasis (z) \, \dd S
\end{align}
is the \emph{solution vector}.
If we can find the general solution to the integral in Equation (\ref{eq:solution_vector}),
we can compute the occultation flux for arbitrary order limb darkening.
The solutions for the case of uniform ($\mathfrak{s}_0$), linear ($\mathfrak{s}_1$),
and quadratic ($\mathfrak{s}_2$) limb darkening have been studied in
the past, so we dedicate sections \S\ref{sec:uniform}--\ref{sec:quadratic}
to revisiting existing formulae and algorithms for computing them, with both speed and numerical accuracy
in mind. The subsequent section (\S\ref{sec:higher_order}) tackles the case of higher order
limb darkening.

\subsection{Normalization}
\label{sec:normalization}
Before we discuss how to compute $\mathfrak{s}$, we turn our attention to
the normalization constant $I_0$. It is convenient to
choose a normalization such that the total unocculted flux is unity (for some choice of units), 
regardless of the value of the limb darkening coefficients. We therefore require that
\begin{align}
    \label{eq:normalization1}
    F &= \iint I(z) \, \dd S \nonumber \\
      &= I_0 \, \mathfrak{s}^\mathsf{T} \mathcal{A} \mathbf{u} \nonumber \\
      &= 1
\end{align}
when the integral is taken over the entire disk of the body. We must thus have
\begin{align}
    \label{eq:normalization2}
    I_0 &= \frac{1}{\mathfrak{s}^\mathsf{T}_{r=0} \mathcal{A} \mathbf{u}} \nonumber \\[0.5em]
        &= \frac{1}{\mathfrak{s}^\mathsf{T}_{r=0} \mathfrak{g}} \quad,
\end{align}
where $\mathfrak{s}^\mathsf{T}_{r=0}$ is the solution vector when there is no
occultor (i.e., $r = 0$). When there is no occultor, the $n^\mathrm{th}$ term of $\mathfrak{s}$
corresponds to the double integral in polar coordinates
\begin{align}
    \label{eq:normalization3}
    \mathfrak{s}_{n,r=0} &= \int_0^{2\pi}\int_0^1 \gbasisn(z) \, r^\prime \, \dd r^\prime \, \dd\theta \nonumber \\[0.5em]
                         &= 2\pi \int_0^1 \gbasisn(z) \, r^\prime \, \dd r^\prime \quad,
\end{align}
From Equation~(\ref{eq:greensbasis}), we may write
\begin{align}
    \label{eq:normalization4}
    \mathfrak{s}_{n,r=0} &=
    2\pi
    \begin{dcases}
        \int_0^1 r \, \dd r & \qquad n = 0
        \\
        \int_0^1 z \, r \, \dd r & \qquad n = 1
        \\
        (n+2) \int_0^1 z^n \, r \, \dd r
        - n \int_0^1 z^{n-2} \, r \, \dd r
        & \qquad n \ge 2 \quad.
    \end{dcases}
\end{align}
The first case is trivial and integrates to $\mathfrak{s}_{0,r = 0} = \pi$.
The remaining cases involve integrands of the form $z^n r$,
where $z = \sqrt{1 - r^2}$. We may evaluate
these integrals by substituting $u = z^2 = 1 - r^2$ and $\dd u = -2r \, \dd r$:
\begin{align}
    \label{eq:normalization5}
    \int_0^1 z^n \, r \, \dd r &= \frac{1}{2} \int_0^1 u^\frac{n}{2} \, \dd u \nonumber \\[0.5em]
                               &= \frac{1}{2 + n} \quad.
\end{align}
The solution vector then simplifies to
\begin{align}
    \label{eq:normalization6}
    \mathfrak{s}_{n,r=0} &=
    \begin{dcases}
        \pi & \qquad n = 0
        \\
        \frac{2\pi}{3} & \qquad n = 1
        \\
        0 & \qquad n \ge 2 \quad.
    \end{dcases}
\end{align}
Interestingly, the net flux contribution for all terms in the Green's basis with
$n \ge 2$ is exactly zero. We may finally evaluate our normalization constant:
\begin{eqnarray}
    \label{eq:normalization}
    I_0 &=& \frac{1}{\pi(\mathfrak{g}_0+ \tfrac{2}{3} \mathfrak{g}_1)}.
\end{eqnarray}

\section{Uniform brightness}

\label{sec:uniform}

Evaluation of the transit light curve of a uniformly bright star, $I(\upmu)=1$, 
amounts to computing the
area of overlap of two disks \citep{MandelAgol2002}.  This has a well-known
analytic solution \citep[e.g.][]{Weisstein2018};  however, we find that the
standard formula leads to round-off error that is larger than necessary
or desirable.  In this section we present a new formula which we demonstrate
yields {\edited double} precision for the area of overlap, along with its derivatives.

\begin{figure}[t!]
    \begin{centering}
    \includegraphics[width=\linewidth]{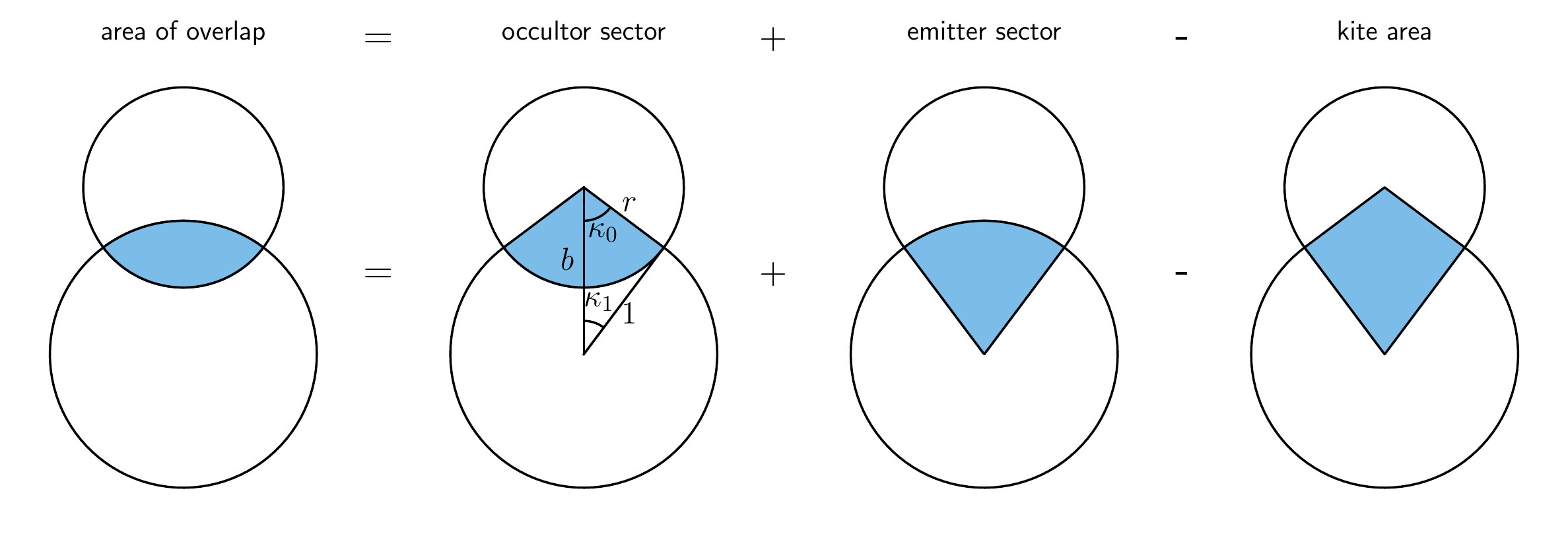}
    \caption{The area of overlap of two circles can be computed as the sum of
    the area of the sectors formed by the centers of each circle and the
    boundary between the points of intersection, minus the area of the kite-shaped
    region formed by the centers of the circles and the intersection points.
    \pycodelink{overlap}\label{fig:circle_overlap}}
    \end{centering}
\end{figure}

Figure \ref{fig:circle_overlap} shows how the area of overlap can be computed
for two circles.  The sums of the areas of the sectors of each circle which span
the area of overlap, minus the area of a kite-shaped region which connects the
centers of the circles with their points of intersection gives the area of the
lens-shaped region of overlap of the two circles.

Taking the radius of the larger circle to be unity, the standard formula for
the lens-shaped overlap area is given by
\begin{proof}{lens_area} \label{eq:MAuniform}
A_{lens} = \pi \Lambda^e(r,b) &=
\begin{dcases}
0 & \qquad 1+r \le b,\\
r^2 \kappa_0 + \kappa_1 -\sqrt{\frac{4b^2-(1+b^2-r^2)^2}{4}} & \qquad \vert 1-r\vert < b < 1+r,\\
\pi r^2 & \qquad b \le 1-r,\\
\pi & \qquad b \le r-1,\\
\end{dcases}
\end{proof}
\citep[e.g.][]{MandelAgol2002}, where
\begin{eqnarray}\label{eq:cosine_formulation}
\cos{\kappa_0} &=& \left(\frac{(r-1)(r+1)+b^2}{2br}\right),\nonumber\\
\cos{\kappa_1} &=& \left(\frac{(1-r)(1+r)+b^2}{2b}\right),
\end{eqnarray}
and $\kappa_0$ and $\kappa_1$ are the angles defined in Figure \ref{fig:circle_overlap}.
The second term in Equation~(\ref{eq:MAuniform}) is the same as the standard formula for the area of overlap of two
partially overlapping circles, with one of the circles scaled to a radius of unity
\citep{Weisstein2018}.  This term corresponds to ingress (and egress), it is the most
expensive to compute, and it is most subject to numerical inaccuracy;  we focus on this
term in what follows.

We find that numerical round-off error limits the precision of the ingress formula when
$b \approx 0$, $b+r \approx 1$, or $b \approx 1+r$;  these are the cases in which
the kite-shaped region becomes thin, in which the sum of two sides becomes similar
in length to the spine of the kite.  The square root term in this formula (Equation
\ref{eq:MAuniform}) computes the area of the kite-shaped region, which in this
form causes round-off error when the kite is flattened.  The same issue occurs when
computing the area of a triangle in which two of the sides are of similar length;
the kite has an area that is twice the area of the {\edited two mirror-image triangles}
connecting the centers
of both circles and one of the intersection points.  \cite{Goldberg1991} gives a
formula for precisely computing the area of a triangle, based on a method developed
by William Kahan \citep[later described in][]{Kahan2000}, which we use to compute
the area of the kite-shaped region,
\begin{eqnarray}\label{eq:Kite_area}
A_{kite} &=& \frac{1}{2}\sqrt{(A+(B+C))(C-(A-B))(C+(A-B))(A+(B-C))},
\end{eqnarray}
for $A \ge B \ge C$, where the tuple $\{A,B,C\}$ equals $\{1,r,b\}$
sorted from from greatest to least.  Note that the order of operations
needs to be carried out as specified by the series of parentheses in
the entry to the square root;  this sequence of operations preserves
numerical precision. {\edited This formula is a novel implementation of Heron's
formula for a triangle for which loss of precision occurs due to subtracting 
quantities with similar numerical values which differ at high significant digits,
and thus are more subject to round-off errors.}

Next, the inverse cosine formulae are also imprecise when $\cos{\kappa_0} = x_0 \approx
1$ or $\cos{\kappa_1} = x_1 \approx 1$.  The approximate solutions in this limit
are $\kappa_0 \approx [2(1-x_0)]^{1/2}$ and $\kappa_1 \approx [2(1-x_1)]^{1/2}$, and so round-off
error can occur both in taking the difference of two numbers close to unity,
and in taking the square root.

Instead, we use the function $\theta = \mathrm{atan2}(y,x)$ with $y=\sin{\theta}$ and
$x=\cos{\theta}$ to compute $\kappa_0$ and $\kappa_1$, which avoids the quadrant and
division-by-zero problems of the $\theta = \tan^{-1}(y/x)$ function.  In addition
to the cosine values above, we require the sine terms, which are given by
\begin{eqnarray}
\sin{\kappa_0} &=& \frac{A_{kite}}{br},\nonumber\\
\sin{\kappa_1} &=& \frac{A_{kite}}{b},
\end{eqnarray}
which can be derived from the area of the triangles formed by the centers of
the circles and one intersection point.
Note that both $\sin{\kappa_0}$ and $\cos{\kappa_0}$ are divided by $br$, and
$\sin{\kappa_1}$ and $\cos{\kappa_1}$ are divided by $b$, so that
in the arctangent formula these denoninator terms cancel, which can improve
numerical stability for small values of $b$ or $r$; this cancellation does not happen
in the arccosine case given in Equation~(\ref{eq:cosine_formulation}).

This results in the following equations for the overlap area, $A_{lens}$, of two
partially overlapping circles:
\begin{proof}{kite_area} \label{eq:area_of_overlap}
A_{lens} &= \kappa_1 + r^2\kappa_0 - A_{kite},\nonumber\\
\kappa_0 &= \mathrm{atan2}(2A_{kite},(r-1)(r+1)+b^2),\nonumber\\
\kappa_1 &= \mathrm{atan2}(2A_{kite},(1-r)(1+r)+b^2),
\end{proof}
with $A_{kite}$ given in Equation~(\ref{eq:Kite_area}).

\begin{figure}[t!]
    \begin{centering}
    \includegraphics[width=\linewidth]{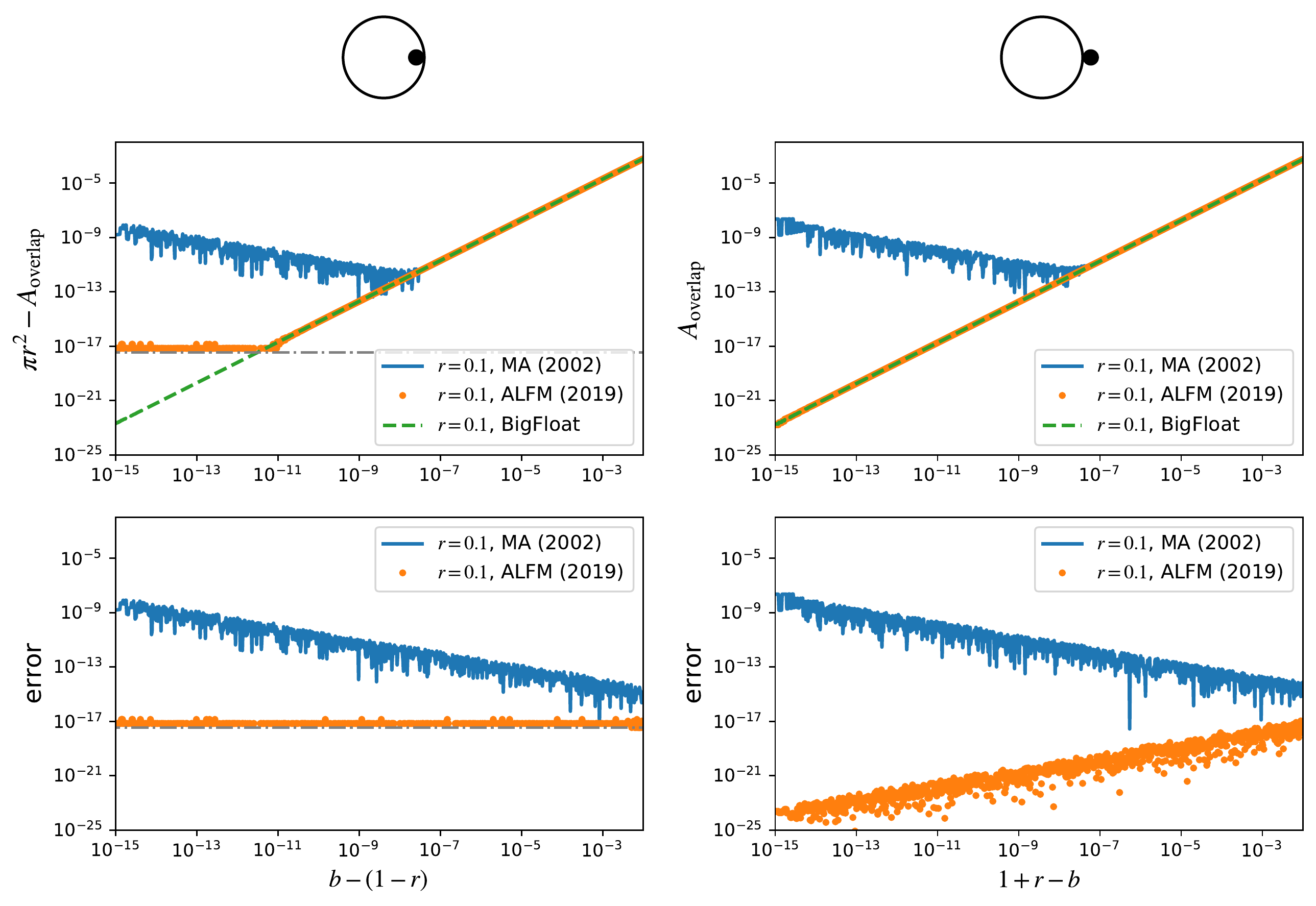}
    \caption{Precision of formulae for the area of overlap of two circles with
    radius ratio $r$.  Plotted are the regions near $b=1-r$ (second and third
    points of contact) and $b=1+r$ (first and fourth points of contact) for
    the standard formula (Equation \ref{eq:MAuniform}, blue) and our new formula
    (Equation \ref{eq:area_of_overlap}, orange dots).
    The high-precision calculation is shown in green dashed for comparison; this
    is limited by the conversion of the result to double-precision. The
    solid and empty circles (top) indicate the positions of the circles at the left
    hand side of the axes.  In the left panels the horizontal dash-dotted grey line
    indicates the limiting precision for representing $\pi r^2$. \jlcodelink{area_of_overlap_r01}\label{fig:overlap_precision}}
    \end{centering}
\end{figure}

The performance of this formula relative to the standard formula is profiled
in Figure \ref{fig:overlap_precision} for $r=0.1$, a typical
value for transiting exoplanets.  We have carried out
the computation in the \texttt{Julia} language, both in double-precision
(\texttt{Float64}), and 256-bit precision (\texttt{BigFloat}), and
subtracted the results to measure the numerical errors of the computation.

We find that the standard
formula (Equation \ref{eq:MAuniform}) approaches errors of $10^{-8}$
in the limit of $b \rightarrow 1-r$. This error exceeds the value
of the area of the smaller circle minus the area of overlap for
values of $1-r < b < 1-r+10^{-8}$.  Thus, even though this calculation
is carried out in double precision, the precision achieved is of
order single precision.  Likewise, for $b \rightarrow 1+r$,
the error of the standard formula approaches $10^{-8}$, with the
error exceeding the value of the area of overlap for $1+r-10^{-8} <
b < 1+r$.

In contrast, Equation~(\ref{eq:area_of_overlap}) gives a precision
that is double-precision in both limits.  Figure
\ref{fig:overlap_precision} shows that Equation~(\ref{eq:area_of_overlap})
gives a precision of $\approx 10^{-17}$ in the limit $b \rightarrow
1-r$ for $r=0.1$; this limit is due to the limiting precision of
representing $\pi r^2$ in double-precision, which in this case
is $\pi r^2 / 2^{53} = 10^{-17.4}$, indicated with a dash-dot
grey line in the left hand panels of Figure \ref{fig:overlap_precision}.
At the beginning of ingress/end of egress when $b \approx 1+r$,
even higher precision is achieved since the area of overlap approaches
zero, as shown in the right hand panels of Figure \ref{fig:overlap_precision}.

Finally, we compute the corresponding element of the solution vector $\mathfrak{s}_0$
as
\begin{eqnarray} \label{eq:uniform}
    \label{eq:s0}
    \mathfrak{s}_0(r,b) &=& \pi-A_{lens}\nonumber\\
                        &=&\pi -\kappa_1 - r^2\kappa_0 + A_{kite}.
\end{eqnarray}
We note that instead of computing $\kappa_1$,
we compute $\pi-\kappa_1 = -\mathrm{atan2}(\sin{\kappa_1},\cos{\kappa_1})$,
which leads to {\edited double} precision as well. Note also that $\mathfrak{s}_0$ is
identical to the first basis function ($s_0$) in the \starry implementation
from \citet{starry}.

\subsection{Derivatives}

The partial derivatives of this formula with respect to the radius
ratio, $r$, and impact parameter, $b$, turn out to be straightforward:
\begin{proof}{dS0drb}\label{eq:dS0_drb}
\frac{\partial \mathfrak{s}_0(r,b)}{\partial r} &= -2r \kappa_0,\nonumber\\
\frac{\partial \mathfrak{s}_0(r,b)}{\partial b} &= \frac{2A_{kite}}{b},
\end{proof}
which can be computed from the quantities already used in calculating $\mathfrak{s}_0$.
At the contact points, when $b = \vert 1\pm r\vert$, the derivatives are undefined.
In practice this can be a problen when taking finite-differences across
the discontinuous boundary, but with the analytic formulae, these points
are a set of measure zero, and so we simply set the derivatives to zero
at these points.

In the remainder of this paper we will need to use these formulae in computing
the higher order limb-darkened light curves.  In the next section, we revisit the formulae
for linear limb darkening.

\section{Linear limb darkening}
\label{sec:reparam}

We now turn to the case of linear limb darkening, $I(\upmu)/I_0 = 1-u_1(1-\upmu)$
\citep{Russell1912a,Russell1912b}.  In this section we set $I_0=1$ and $u_1=1$, so
that $I(\upmu)=\upmu$, which corresponds to the $n=1$ terms in both the polynomial
and Green's bases;  the general linear limb-darkening case can be computed as a linear
combination with the uniform case.
Note that since $\upmu = \sqrt{1-x^2-y^2}$, this problem is equivalent to
computing the volume of intersection between a sphere and a cylinder, which was
solved in terms of elliptic integrals by \citet{Lamarche1990}.
A similar solution was found by \citet{MandelAgol2002}, who show that the total
flux visible during the occultation of a body whose surface map is given by
$I(x, y) = \sqrt{1 - \x^2 -\y^2}$ may be computed as
\begin{align}
    \label{eq:s1}
    \mathfrak{s}_1 = \frac{2\pi}{3} \left(1 - \frac{3\Lambda(r,b)}{2} - \Theta(r - b) \right)
\end{align}
where $\Theta(\bigdot)$ is the Heaviside step function and
\begingroup\makeatletter\def\f@size{10}\check@mathfonts
\def\maketag@@@#1{\hbox{\m@th\normalsize#1}}%
\begin{proof}{biglam}
    \label{eq:biglam}
    \Lambda(r,b) &=
    \begin{dcases}
          %
          %
          \frac{1}{9 \pi \sqrt{b r}} \Bigg[
                \frac{(r + b)^2 - 1}{r + b}
                \Big(
                    -2r \,
                    \big(
                        2 (r + b)^2 + (r + b)(r - b) - 3
                    \big)
                    K(k^2)
                    &\\ \phantom{XXXX}
                    + 3 (b - r) \, \Pi\big(k^2 (b + r)^2, \, k^2\big)
                \Big)
                - 4 b r (4 - 7 r^2 - b^2) E(k^2)
          \Bigg]
          & \qquad k^2 < 1
          \\[1.5em]
          \frac{2}{9 \pi} \Bigg[
                \big(1 - (r + b)^2\big)
                \Bigg(
                    \sqrt{1 - (b - r)^2} \,
                    K\left(\frac{1}{k^2}\right)
                    + 3 \left(\frac{b-r}{(b+r)\sqrt{1 - (b - r)^2}}\right)
                    &\\ \phantom{XX}
                    \times \Pi\left(\frac{1}{k^2(b+r)^2}, \, \frac{1}{k^2}\right)
                \Bigg)
                - \sqrt{1 - (b - r)^2}
                (4 - 7 r^2 - b^2)
                E\left(\frac{1}{k^2}\right)
          \Bigg]
          & \qquad k^2 \ge 1
    \end{dcases}
\end{proof}
\endgroup
with
\begin{align}
    \label{eq:k2}
    k^2 &= \frac{1 - r^2 - b^2 + 2 b r}{4 b r}
    \quad.
\end{align}
Note that $\mathfrak{s}_1(r,b) = s_2(r,b)$ in the spherical harmonic expansion used in \starry as
described in \citet{starry}.
For the cases $b=r$, $b=1-r$, $b=0$, $r=0$, or $\vert r-b\vert \ge 1$, there are special
expressions for $\Lambda(r,b)$ given below.
In the expressions above, $K(\bigdot)$, $E(\bigdot)$, and $\Pi(\bigdot, \bigdot)$
are the complete elliptic integrals of the first, second kind, and third kind,
respectively, defined as
\begin{align}
    \label{eq:elliptic}
    K(k^2) &\equiv \int_0^{\frac{\pi}{2}} \frac{\dd \varphi}{\sqrt{1 - k^2 \sin^2 \varphi}}
    \nonumber \\[0.5em]
    E(k^2) &\equiv \int_0^{\frac{\pi}{2}} \sqrt{1 - k^2 \sin^2 \varphi} \, \dd \varphi
    \nonumber \\[0.5em]
    \Pi(n, k^2) &\equiv \int_0^{\frac{\pi}{2}} \frac{\dd \varphi}{(1 - n \sin^2 \varphi)\sqrt{1 - k^2 \sin^2 \varphi}}
    \quad.
\end{align}
In Equation~(\ref{eq:biglam}) we have transformed the formulae from \citet{MandelAgol2002} using
Equation (17.7.17) from \citet{Abramowitz1970} which yields equations that are better
behaved in the vicinity of $b=r$.\footnote{Note that we corrected several typos
in \citet{MandelAgol2002}, which are listed in the Appendix.}  However, these elliptic
integrals are still subject to numerical instability as $r \rightarrow 1-b$ and $r \gg 1$.
The main issue is the logarithmic divergence of $K$ and $\Pi$ as $k \rightarrow 1$, as
well as numerical cancellations leading to round-off errors which occur in the
limit $k \rightarrow 0$.

Through trial and error, we have found that these instabilities can be removed by combining
elliptic integrals into a general complete elliptic integral defined by \citet{Bulirsch1969} as
\begin{equation}\label{eq:cel}
\mathrm{cel}(k_c,p,a,b) = \int_0^{\pi/2} \frac{a\cos^2{\phi} + b\sin^2{\phi}}{\cos^2{\phi}+p\sin^2{\phi}} \frac{d\phi}{\sqrt{\cos^2{\phi}+k_c^2\sin^2{\phi}}},
\end{equation}
where $k_c = \sqrt{1-m_k}$, and for $b+r \ge 1$,
$m_k=k^2$, while for $b+r \le 1$, $m_k=1/k^2$.  The derivatives of
$\mathrm{cel}$ with respect to the input parameters are given in Appendix \ref{app:cel_derivatives}.
Although $k_c$ can be computed from
$m_k$, we have found better numerical stability in computing $k_c$ analytically
from $b$ and $r$:
\begin{align}
    k_c &=
    \begin{dcases}
     \sqrt{\frac{(b+r)^2-1}{4br}} & \qquad k^2 \le 1\\
     \sqrt{\frac{1-(b+r)^2}{1-(b-r)^2}} & \qquad k^2 > 1.
   \end{dcases}
\end{align}
In practice, we let the subroutine that computes $\mathrm{cel}$ accept both
$m_k$ and $k_c$ as input for numerical precision.

To transform the elliptic integrals in Equation~(\ref{eq:biglam}) to $\mathrm{cel}$,
we used the following relations from \citet{Bulirsch1969}:
\begin{eqnarray} \label{eq:cel_identities}
\lambda K(m_k) + q E(m_k) &=& {\rm cel}(k_c,1,\lambda+q,\lambda+q k_c^2)\\
\lambda K(m_k) + q \Pi(n,m_k) &=& {\rm cel}(k_c,1-n,\lambda+q,\lambda (1-n) + q)\\
E(m_k) &=& {\rm cel}(k_c,1,1,1-m_k)\\
E(m_k)-(1-m_k)K(m_k) &=& m_k \, {\rm cel}(k_c,1,1,0)\\
\Pi(n,m_k)-K(m_k)  &=& n \, {\rm cel}(k_c,1-n,0,1),
\end{eqnarray}
noting that \citet{Bulirsch1969} uses a different sign convention for $\Pi(n,m_k)$.
In particular, the expressions for $\Pi(n,m_k)-K(m_k)$ and $E(m_k)-(1-m_k)K(m_k)$ are useful for eliminating
the singularities and cancellations which occur at $m_k=1$ when $b+r=1$ and $m_k=0$ when
$r \rightarrow \infty$.  The general complete elliptic integral is evaluated
with the approach of \citet{Bartky1938}, which uses recursion to approximate the
integral to a specified precision.

These elliptic integral transformations lead to the following numerically-stable
expression for the linear limb darkening flux, $\mathfrak{s}_1(r,b)$, in which
\begin{proof}{biglam_stable}
    \label{eq:biglam_stable}
    \Lambda &=
    \begin{dcases}
          0 & \qquad  r = 0\\
          0 & \qquad  \vert r- b\vert \ge 1\\
          -\tfrac{2}{3}(1-r^2)^{3/2} & \qquad b = 0\\
          \tfrac{1}{3} - \tfrac{4}{9\pi} & \qquad b = r = \tfrac{1}{2}\\
          \tfrac{1}{3} + \tfrac{2}{9\pi} {\rm cel}\left(k_c,1,m_k-3,(1-m_k)(2m_k-3)\right) & \qquad b= r < \tfrac{1}{2}\\
          \tfrac{1}{3} + \tfrac{4r}{9\pi} {\rm cel}\left(k_c,1,1-3m_k,m_k-1\right) & \qquad b= r > \tfrac{1}{2}\\  
          \tfrac{2}{9\pi}\left[3\cos^{-1}(1-2r) -2(3+2r-8r^2)\sqrt{rb}-3\pi\Theta(r-\tfrac{1}{2})\right] & \qquad b+r =1\\
          \frac{1-(b-r)^2}{9 \pi \sqrt{b r}} \Bigg[
                \frac{(b+r)^2-1}{4br}(b^2-r^2){\rm cel}(k_c,(b-r)^2(1-m_k),0,3)
                &\\ \phantom{XXXX}
               - (3-6r^2-2br){\rm cel}(k_c,1,1,0)-4brE(m_k)
          \Bigg]
          & \qquad k^2 < 1
          \\[1.5em]
          \frac{2\sqrt{1-(b-r)^2}}{9 \pi} \Bigg[
                \big(1 - (r + b)^2\big)
                {\rm cel}(k_c,p,1+q,p+q) &\\ \phantom{XXXX}
                - (4 - 7 r^2 - b^2)
                E\left(m_k\right)
          \Bigg]
          & \qquad k^2 > 1\\
    \end{dcases}
\end{proof}
where
\begin{eqnarray}
q &=& 3\frac{b-r}{(b+r)(1-(b-r)^2)}\nonumber\\
p &=& \left(\frac{b-r}{b+r}\right)^2 \frac{1-(b+r)^2}{1-(b-r)^2}
\end{eqnarray}
in the $k^2 > 1$ case.  Note that in this equation the conditions
should be evaluated in the order they appear.

\begin{figure}[p!]
    \begin{centering}
    \includegraphics[height=3in]{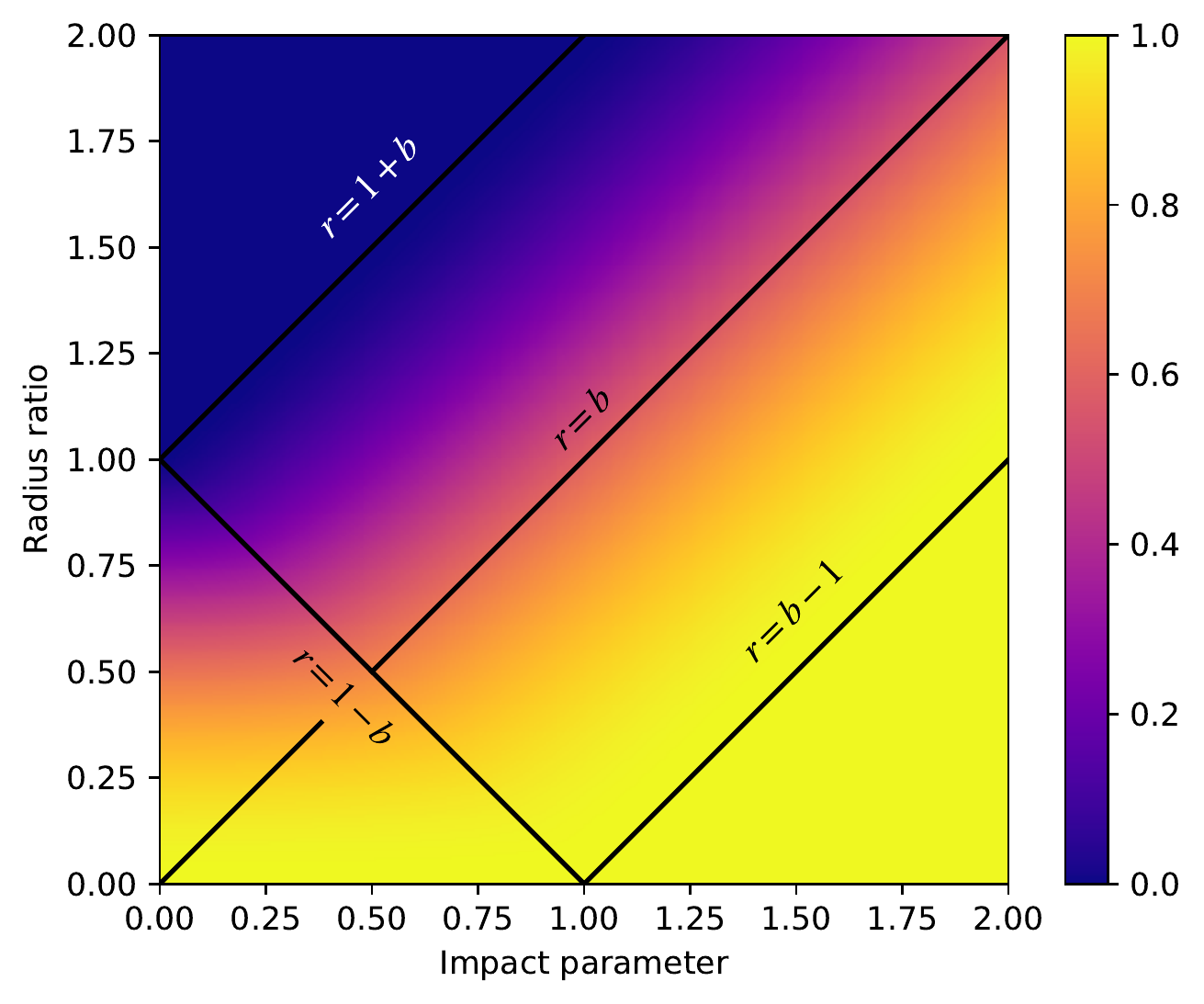}
    \caption{{\edited The flux of a linearly limb-darkened star being eclipsed, 
    $F = \tfrac{3}{2\pi} \mathfrak{s}_1(r,b)$, with $u_1=1$ (all other $u_n$ zero), 
    for which $I_0 = 3/(2\pi)$.
    In the limit $b > r+1$, no eclipse occurs, so $F=1$.  For $b < r-1$, the star
    is completely eclipsed and $F=0$.  In the limits $b=r$ and $b=1-r$, special
    expressions must be used.}
    \jlcodelink{transit_linear}\label{transit_linear}}
    \end{centering}
\end{figure}

\begin{figure}[p!]
    \begin{centering}
    \includegraphics[height=4in]{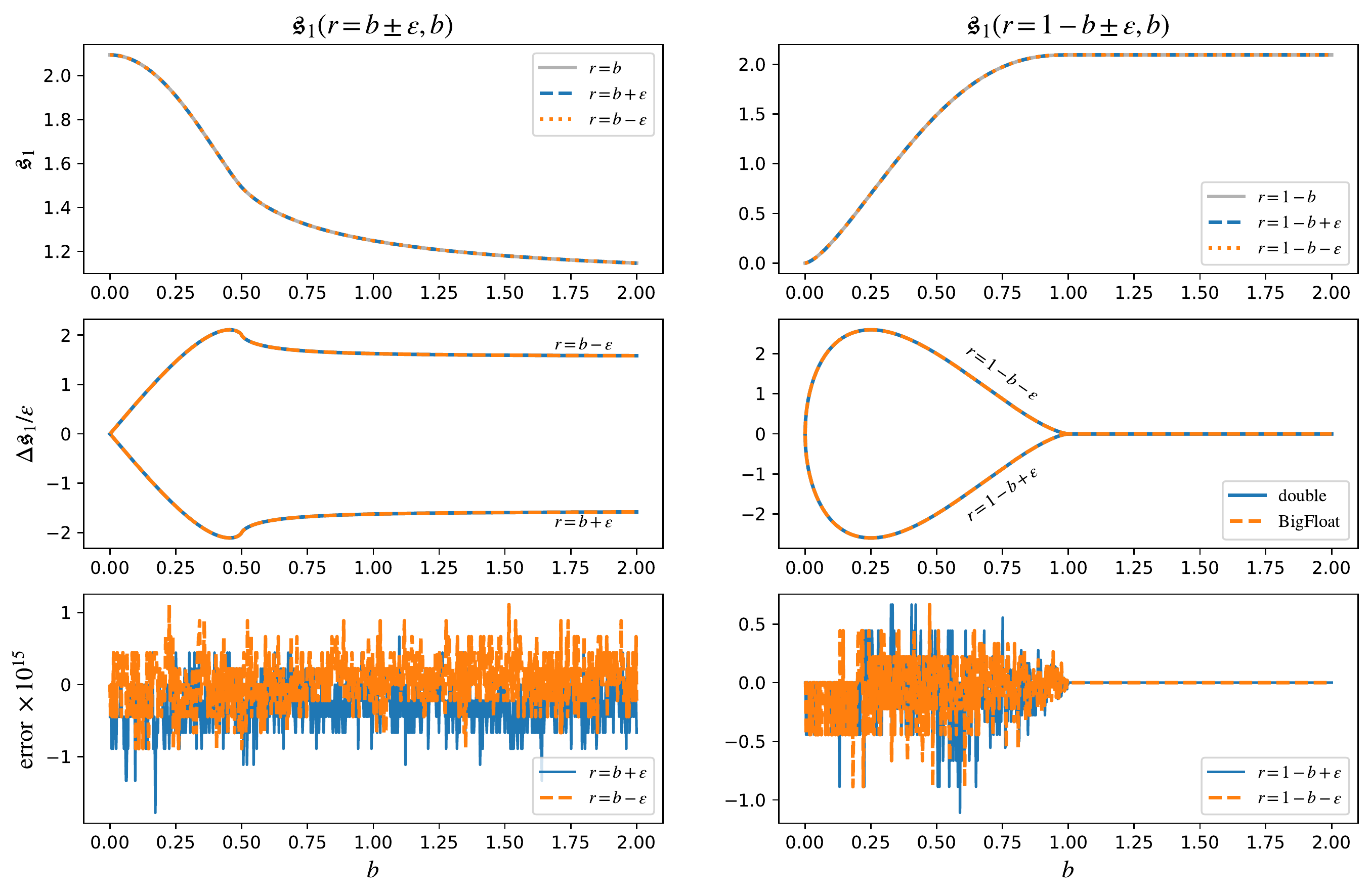}
    \caption{The accuracy of $\mathfrak{s}_1(r,b)$ near $b=r$ (left panels) and
    $b=1-r$ (right panels) for $\epsilon = 10^{-8}$. The $x$-axes are impact parameter $b$,
    while the $y$ axes in the top panels show $\mathfrak{s}_1(r,b)$, with $r$
    given in the legend of each panel. The middle panels plot
    the difference $(\mathfrak{s}_1(b\pm\epsilon,b)-\mathfrak{s}_1(b,b))/\epsilon$
    and $(\mathfrak{s}_1(1-b\pm\epsilon,b)-\mathfrak{s}_1(1-b,b))/\epsilon$. The bottom
    panels show the numerical precision by the comparing double precision
    computation with \texttt{BigFloat} precision (256-bit). \jlcodelink{s2_machine}\label{s2_machine}}
    \end{centering}
\end{figure}

The $\mathfrak{s}_1(r,b)$ function is plotted in Figure \ref{transit_linear}. The
function varies smoothly from the lower right where the disk is
unocculted to the upper left where it is completely occulted.
There are several points which need to be handled separately as
the Equation~(\ref{eq:biglam}) expressions become singular or are
no longer valid;  the solid lines in Figure \ref{transit_linear} show
these points.  When $b=0$, the integral over the center of the
disk simplifies greatly.  When $b=r=1/2$, at the intersection of
$b=r$ and $b=1-r$, another simplification occurs.  For $b=r$,
the disk of the occultor crosses the center of the source;
this needs to be computed separately in the $r<1/2$, $r=1/2$,
and $r>1/2$ limits.  The first and fourth contacts occur at
$b=1+r$, where $\mathfrak{s}_1=1$;  this is the upper bound to the $k^2 < 1$
region for $b+r >1$.
For $r \ge 1$, the second and third contacts (at the start and
end of complete occultation) occur when $b=1-r$, which is the
lower  bound to the $k^2<1$ region when $b+r >1$.
For $r < 1$, the second and third contacts occur when $r=1-b$.

Near these boundaries, the standard \citet{MandelAgol2002} expressions
can become singular, and so we paid particular care to the accuracy of these
new expressions in these regions.  Figure \ref{s2_machine} shows
that Equation~(\ref{eq:biglam_stable}) is accurate to double
precision in all of these regimes.
We tested the accuracy by computing the equations with 256 bit
arithmetic, which is much less subject to round-off error, and
hence gives more precise expressions than double precision.  We implemented the
pseudocode from \citet{Bulirsch1969} to compute ${\rm cel}(k_c,p,a,b)$,
which has a termination test that scales as the square root of
the {\edited double} precision.  We find that the transformed expressions
are accurate to $\sim 10^{-14}$ when computed in double precision
within $\epsilon = 10^{-8}$ of the vicinity of $b=r$ and $b=1-r$.

\begin{figure}[p!]
    \begin{centering}
    \includegraphics[width=0.8\linewidth]{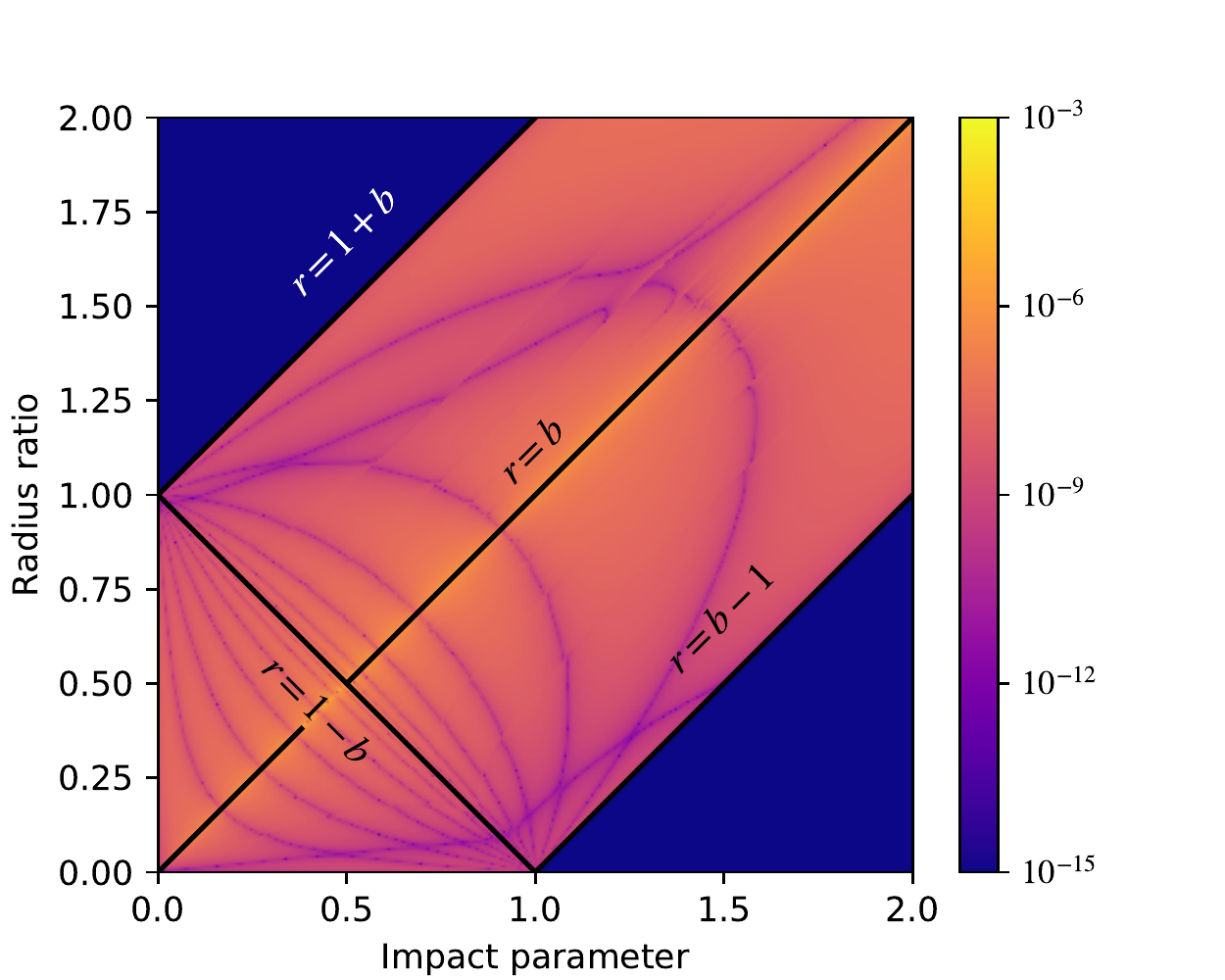}
    \caption{The numerical error in computing the flux of an eclipsed, linearly
             limb-darkened star ($u_1=1$) using the equations in \citet{MandelAgol2002}.
             \jlcodelink{s2_residuals_MA2002}\label{fig:s2_plot_MA2002}}
    \end{centering}
\end{figure}

\begin{figure}[p!]
    \begin{centering}
    \includegraphics[width=0.8\linewidth]{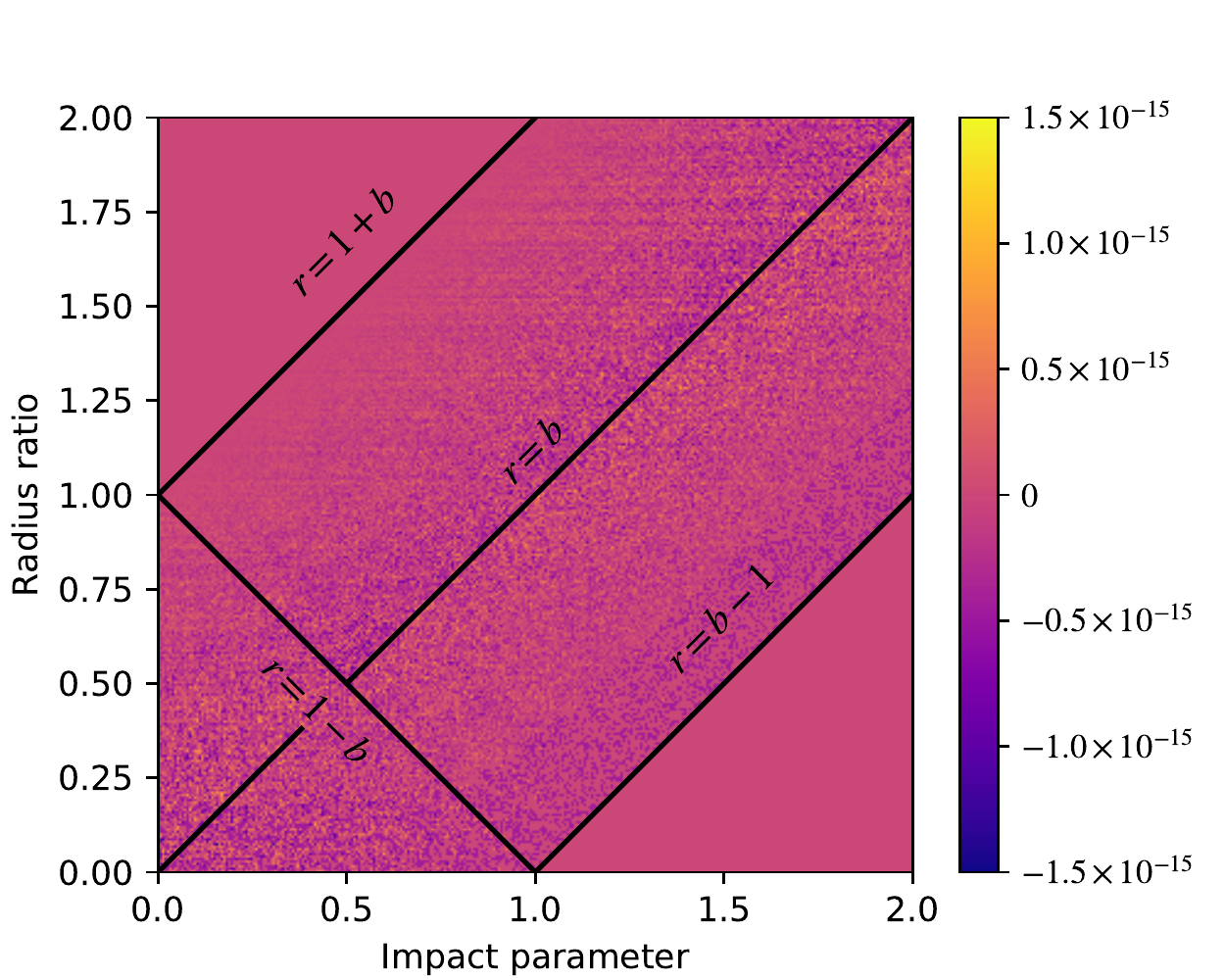}
    \caption{The numerical error in computing the flux of an eclipsed, linearly
    limb-darkened star ($u_1=1$) using the $\mathfrak{s}_1(r,b)$ formalism introduced in this
    paper. Compare to Figure~\ref{fig:s2_plot_MA2002}, noting the change in the color
    scale. The new method is eight orders of magnitude more precise on average,
    approaching double precision accuracy everywhere in the domain.
    \jlcodelink{s2_residuals}\label{fig:s2_plot}}
    \end{centering}
\end{figure}

Finally, in Figures~\ref{fig:s2_plot_MA2002} and \ref{fig:s2_plot} we plot
the relative numerical error in the flux of a linearly limb-darkened source
when using the equations in \citet{MandelAgol2002} and in this paper,
respectively, over a portion of the $b-r$ plane. The former method
(Figure~\ref{fig:s2_plot_MA2002}) yields errors
on the order of $10^{-7}$ over most of the domain, although the error approaches
unity near the singular regions discussed above. In contrast, the method introduced
in this paper (Figure~\ref{fig:s2_plot}; note the change in the color scale)
yields errors close to {\edited double} precision everywhere, including the vicinity of the
singular points.

\pagebreak 

\subsection{Derivatives}%
The derivatives of $\Lambda$ with respect to $r$ and $b$ are:
\begingroup\makeatletter\def\f@size{10}\check@mathfonts
\def\maketag@@@#1{\hbox{\m@th\normalsize#1}}%
\begin{proof}{biglam_deriv}
    \label{eq:dbiglam_dr}
    \frac{\partial \Lambda}{\partial r} &=
    \begin{dcases}
          0
          \phantom{MMMMMMMMMMMMMMMMMMM} 
          & \qquad  r = 0\\
          0 & \qquad  \vert r- b\vert \ge 1\\
          2 r\sqrt{1-r^2} & \qquad b = 0\\
           \frac{2}{\pi} & \qquad b = r = \tfrac{1}{2}\\
          \frac{4r}{\pi} E(4r^2) & \qquad b= r < \tfrac{1}{2}\\
          \frac{2}{\pi} \mathrm{cel}(k_c,1,1,0) & \qquad b= r > \tfrac{1}{2}\\
          \frac{8r}{\pi}\sqrt{r(1-r)} & \qquad b+r =1\\
          \frac{8br^2 E(k^2) + 2r(1-(b+r)^2)K(k^2)}{\pi\sqrt{br}}
                    &\\ \phantom{XX}
          = \frac{1}{\pi\sqrt{br}}\mathrm{cel}(k_c,1,2r(1-(b-r)^2),0) & \qquad k^2 < 1
          \\[1.5em]
          \frac{4r}{\pi}\sqrt{1-(b-r)^2} E(k^{-2})
                    &\\ \phantom{XX}
          = \frac{4r}{\pi}\sqrt{1-(b-r)^2} \mathrm{cel}(k_c,1,1,k_c^2)& \qquad k^2 > 1
          \\
    \end{dcases}
\end{proof}
\endgroup
and
\begingroup\makeatletter\def\f@size{10}\check@mathfonts
\def\maketag@@@#1{\hbox{\m@th\normalsize#1}}%
\begin{proof}{biglam_deriv}
    \label{eq:dbiglam_db}
    \frac{\partial \Lambda}{\partial b} &=
    \begin{dcases}
          0 & \qquad  r = 0\\
          0 & \qquad  \vert r- b\vert \ge 1\\
          0 & \qquad b = 0\\
           -\frac{2}{3\pi} & \qquad b = r = \tfrac{1}{2}\\
          \frac{4r}{3\pi}\mathrm{cel}(k_c,1,-1,k_c^2) & \qquad b= r < \tfrac{1}{2}\\
          -\frac{2}{3\pi} \mathrm{cel}(k_c,1,1,2k_c^2) & \qquad b= r > \tfrac{1}{2}\\
          -\frac{8r}{3\pi}\sqrt{r(1-r)} & \qquad b+r =1\\
           \frac{4r(r^2+b^2-1) E(k^2) + 2r(1-(b+r)^2)K(k^2)}{3\pi\sqrt{br}}
                    &\\ \phantom{XX}
          = \frac{1-(b-r)^2}{3\pi \sqrt{br}} \mathrm{cel}(k_c,1,-2r,(1-(b+r)^2)/b) & \qquad k^2 < 1
          \\[1.5em]
          \frac{2}{3b\pi}\sqrt{1-(b-r)^2}\left[(r^2+b^2-1) E(k^{-2}) +(1-(b+r)^2)K(k^{-2})\right]
                    &\\ \phantom{XX}
          =\frac{4r}{3\pi}\sqrt{1-(b-r)^2}\mathrm{cel}(k_c,1,-1,k_c^2) & \qquad k^2 > 1,\\
    \end{dcases}
\end{proof}
\endgroup
where we have given some of the expressions in terms of both the standard elliptic integrals
and the general elliptic integral.


From these expressions, the derivatives of $\mathfrak{s}_1$ are given by
\begin{eqnarray}
\frac{\partial \mathfrak{s}_1}{\partial r} &=& -\pi \frac{\partial \Lambda}{\partial r},\\
\frac{\partial \mathfrak{s}_1}{\partial b} &=& -\pi \frac{\partial \Lambda}{\partial b}.
\end{eqnarray}

We have tested these formulae with finite-difference derivatives evaluated at
256-bit precision, and, as with the total flux term, we find that these are accurate
to $\la 2 \times 10^{-15}$, close to {\edited double} precision.

We next increase the power of limb darkening by one, $\upmu^2$.

\section{Quadratic limb darkening}
\label{sec:quadratic}

The next order of limb darkening has been widely studied due to its
accurate description of stellar atmospheres \citep{Claret2000,MandelAgol2002,Pal2008}.
We summarize here the formulae for quadratic limb darkening for $I(\upmu)=4\upmu^2-2$%
, along with the derivatives, using the transformed
expressions described above. The general quadratic case may be computed as
a linear combination with the foregoing uniform and linear cases.

We first give the formula for the function $\eta(r,b)$, which is the term appearing
in the quadratic limb darkening model only when $u_2 \ne 0$
\citep{MandelAgol2002}. In terms of quantities we have defined above for the uniform case:
\begin{proof}{Eta}
    \label{eq:eta}
    \eta(r,b) &=
    \begin{dcases}
          \frac{1}{2\pi}\left[\kappa_1+r^2(r^2+2b^2)\kappa_0-\frac{1}{2}(1+5r^2+b^2)A_{kite}\right]
          & \qquad k^2 \le 1
          \\[1.5em]
          \frac{r^2}{2}(r^2+2b^2)
          & \qquad k^2 > 1\\
    \end{dcases}
\end{proof}
As $\kappa_0$, $\kappa_1$, and $A_{kite}$ were already computed in the
uniform case, these quantities are reused in the quadratic computation.

With this definition, the quadratic term, $\mathfrak{s}_2(r,b)$ is given simply by
%
\begin{eqnarray}
    \label{eq:s2}
    \mathfrak{s}_2 &=& 2 \mathfrak{s}_0 + 4\pi \eta - 2\pi \quad,
\end{eqnarray}
where $\mathfrak{s}_0$ is defined in Equation (\ref{eq:uniform}).


\pagebreak 

\subsection{Derivatives}
The derivatives of $\eta$ are given by:
\begin{proof}{Eta}
    \label{eq:detadr}
    \frac{\partial \eta}{\partial r} &=
    \begin{dcases}
          \frac{2r}{\pi}\left[(r^2+b^2)\kappa_0-2A_{kite}\right]
          & \qquad k^2 \le 1
          \\[1.5em]
          2r(r^2+b^2)
          & \qquad k^2 > 1\\
    \end{dcases}
\end{proof}
and
\begin{proof}{Eta}
    \label{eq:detadb}
    \frac{\partial \eta}{\partial b} &=
    \begin{dcases}
          \frac{1}{2b\pi}\left[4r^2b^2\kappa_0-2(1+b^2+r^2)A_{kite}\right]
          & \qquad k^2 \le 1
          \\[1.5em]
          2br^2
          & \qquad k^2 > 1\\
    \end{dcases}
\end{proof}
where the derivatives of $\mathfrak{s}_0$ are defined in Equation (\ref{eq:dS0_drb}).
The derivatives of the $\mathfrak{s}_2$ term are thus
\begin{eqnarray}
    \frac{\partial \mathfrak{s}_2}{\partial r} &=& 2 \frac{\partial \mathfrak{s}_0}{\partial r} + 4\pi \frac{\partial \eta}{\partial r}, \nonumber \\
    \frac{\partial \mathfrak{s}_2}{\partial b} &=& 2 \frac{\partial \mathfrak{s}_0}{\partial b} + 4\pi \frac{\partial \eta}{\partial b} \quad.
\end{eqnarray}

In the following section we turn our attention to the general polynomial limb darkening case,
$\upmu^n$ with $n > 2$.  As discussed in \citet{starry},
these terms may be expressed exactly as the sum of
spherical harmonics with $m=0$. However, it is possible to exploit
the azimuthal symmetry of the limb darkening problem to
derive far more efficient and accurate formulae, which we describe in the following section.

\section{Higher Order Limb Darkening}
\label{sec:higher_order}


Having computed $\mathfrak{s}_0$, $\mathfrak{s}_1$, and $\mathfrak{s}_2$,
we now seek a general expression for $\mathfrak{s}_n$ for any $n > 2$.
Recalling our definition of $\mathfrak{s}$ as the surface integral of
the terms in the Green's basis,
\begin{align}
    \mathfrak{s} \equiv \iint \gbasis (z) \, \dd S \quad,
\end{align}
in this section we will use Green's theorem to re-express this two-dimensional
integral as a one-dimensional line integral over the boundary of the visible
portion of the occulted body's disk. This is the same procedure adopted by
\citet{starry}, albeit with different basis functions.
Given $\bvec{r} = x \xhat + y \yhat$, we may write
\begin{align}
    \label{eq:greens}
    \mathfrak{s} &=
    \oint \bvec{G} (z) \cdot \dd \bvec{r}
    \quad,
\end{align}
where $\bvec{G}$ is a matrix whose $n^{\mathrm{th}}$ row is the
vector
\begin{align}
    \label{eq:greens_n}
    \bvec{G}_n (z) = G_{n,x} (z) \, \xhat + G_{n,y} (z) \, \yhat \quad .
\end{align}
The components $G_{n,x}$ and $G_{n,y}$ are chosen such that
\begin{align}
    \label{eq:DGg}
    \bvec{D} \wedge \bvec{G}_n &\equiv \frac{\dd G_{n,y}}{\dd \x}
                                     - \frac{\dd G_{n,x}}{\dd \y} \nonumber \\
                               &= \gbasisn(z) \quad.
\end{align}
As in \citet{Pal2012} and \citet{starry}, the operation
$\bvec{D} \wedge \bvec{G}_n$ denotes the
\emph{exterior derivative} of $\bvec{G}_n$.
%
%
%
Following \citet{starry},
if we choose the following form for
Equation~(\ref{eq:greens_n}),
\begin{equation}
\mathbf{G}_n(z) = z^n (-y \xhat + x \yhat) \quad,
\end{equation}
we arrive at the expression presented in Equation~(\ref{eq:greensbasis})
for the components of the Green's basis:
\begin{align}
\gbasisn(z)   &= \frac{\dd {G_n}_y}{\dd \x} - \frac{\dd {G_n}_x}{\dd \y} \nonumber \\[0.5em]
              &= (n+2)z^n-n z^{n-2}
\end{align}
for $2 \le n \le N$.
Note that we already introduced the first three terms, $\tilde{\mathfrak{g}}_0 = 1$
(uniform limb darkening), $\tilde{\mathfrak{g}}_1 = z$ (linear limb darkening),
and $\tilde{\mathfrak{g}}_2 = 4z^2 - 2$ (quadratic limb darkening).
Since we already know how to integrate them (\S\ref{sec:uniform}--\ref{sec:quadratic}),
we treat them separately from the higher order terms.


Returning to \eq{greens}, we note that the line integral consists of two arcs:
an arc $\mathcal{P}$ along the boundary of the occulting body and an arc $\mathcal{Q}$ along the
boundary of the occulted body. We may therefore write the $n^\mathrm{th}$ component of
the solution vector as
\begin{align}
    \label{eq:sn}
    \mathfrak{s}_n &= \mathcal{Q}(\bvec{G}_n) - \mathcal{P}(\bvec{G}_n)
    \quad,
\end{align}
where, as in \citet{Pal2012} and \citet{starry}, we define the \emph{primitive integrals}
\begin{align}
    \label{eq:primitivePdef}
    \mathcal{P}(\bvec{G}_n) &\equiv
    \int\displaylimits_{\pi-\phi}^{2\pi + \phi}
        \big[ G_{n,y}(r c_\varphi, b + r s_\varphi) c_\varphi -
              G_{n,x}(r c_\varphi, b + r s_\varphi) s_\varphi \big] r \dd \varphi \quad,
    \\
\intertext{taken along the boundary of the occulting body of radius $r$, and}
    \label{eq:primitiveQdef}
    \mathcal{Q}(\bvec{G}_n) &\equiv
    \int\displaylimits_{\pi-\lambda}^{2\pi + \lambda}
        \big[ G_{n,y}(c_\varphi, s_\varphi) c_\varphi -
              G_{n,x}(c_\varphi, s_\varphi) s_\varphi \big] \dd \varphi
    \quad,
\end{align}
taken along the boundary of the occulted body of radius unity.
For convenience, we defined
$c_\varphi \equiv \cos \varphi$
and
$s_\varphi \equiv \sin \varphi$
and we used the fact that along the arc of a circle,
\begin{align}
    \label{eq:dr}
    \dd \bvec{r} &= -r s_\varphi \, \dd \varphi \, \xhat +
                     r c_\varphi \, \dd \varphi \, \yhat
    \quad.
\end{align}
The angles $\phi$ and $\lambda$ are the same as those used in
\citet{starry} (see their Figure~2) and are given by
$\phi = \kappa_0-\pi/2$ and
$\lambda = \pi/2 - \kappa_1$ (c.f. Equation~\ref{eq:area_of_overlap} and
Figure~\ref{fig:circle_overlap}).

Inserting our expression for $\mathbf{G}_n$ into Equations~(\ref{eq:primitivePdef})
and (\ref{eq:primitiveQdef}), we arrive at a fairly simple form for the primitive
integrals:
\begin{align}
    \label{eq:primitiveP}
    \mathcal{P}(\bvec{G}_n) &=
    \int\displaylimits_{\pi-\phi}^{2\pi + \phi} z^n (r+b \sin{\varphi}) r d\varphi
\intertext{and}
    \label{eq:primitiveQ}
    \mathcal{Q}(\bvec{G}_n) &=
    \int\displaylimits_{\pi-\lambda}^{2\pi + \lambda} z^n d \varphi \quad.
\end{align}

Conveniently, the primitive integral $\mathcal{Q}(\bvec{G}_n) = 0$ for
all $n > 0$, since $z=0$ at the boundary of the star.
Since we need not compute the integral for $n=0$, as we already
found a solution for uniform limb darkening in \S\ref{sec:uniform},
our final task is to find the solution to Equation~(\ref{eq:primitiveP}).

\pagebreak 

\subsection{Solving the $\mathcal{P}$ integral}
\label{sec:Pintegral}

The primitive integral
$\mathcal{P}(\bvec{G}_n)$ can be rewritten as
\begin{equation}
\mathcal{P}(\bvec{G}_n) =
\int_{\pi-\phi}^{2\pi + \phi} \left(1-r^2-b^2-2br s_\varphi\right)^{\frac{n}{2}} (r+b s_\varphi) r d\varphi,
\end{equation}
where $s_\varphi = \sin{\varphi}$.
We make the transformation $\xi = \tfrac{1}{2} \left(\varphi - \tfrac{3\pi}{2}\right)$, yielding
\begin{equation}\label{eq:greens_transformed}
\mathcal{P}(\bvec{G}_n) =
2r (4br)^{\frac{n}{2}}\int\displaylimits_{-\tfrac{\kappa_0}{2}}^{\tfrac{\kappa_0}{2}}
(k^2-\sin^2\xi)^{\tfrac{n}{2}} (r-b + 2b \sin^2 \xi) d\xi,
\end{equation}
for $2 \le n \le N$, where $\kappa_0 = 2 \sin^{-1}k$ for $k^2 \le 1$ and
$\kappa_0 = \pi$ for $k^2 > 1$.  We
reuse the value of $\kappa_0$ which was computed in the uniform limb darkening
case (\S \ref{sec:uniform}).


We can express $\mathcal{P}(\bvec{G}_n)$ in terms of a sequence of integrals,
$\mathcal{M}_n(r,b)$, given by:
\begin{equation}\label{eq:M_of_n}
\mathcal{M}_n(r,b) = (4br)^{n/2} \int_{-\kappa_0/2}^{\kappa_0/2} (k^2-\sin^2\xi)^{\tfrac{n}{2}} d\xi,
\end{equation}
in terms of which the primitive integral takes the particulary simple form
\begin{proof}{pofgn_v01}\label{eq:primitive}
\mathcal{P}(\bvec{G}_n) = (1+r^2-b^2)\mathcal{M}_n - \mathcal{M}_{n+2}.
\end{proof}

The integrals $\mathcal{M}_n$ obey straightforward recursion relations
\begin{proof}{Mn_recursion}\label{eq:Mn_recursion}
\mathcal{M}_n &= \frac{1}{n} \left[ 2(n-1) (1-r^2-b^2) \mathcal{M}_{n-2} \right.\nonumber\\
  &+ \left. (n-2) (1-(b-r)^2)((b+r)^2-1) M_{n-4}\right],\\[1em]
\mathcal{M}_n &= \frac{(n+4)\mathcal{M}_{n+4} - 2(n+3)(1-r^2-b^2)\mathcal{M}_{n+2}}{(n+2)(1-(b-r)^2)((b+r)^2-1)},
\end{proof}
where the first relation may be used for upwards recursion in $n$ for $k^2 > \frac{1}{2}$,
and the second for downward recursion in $n$ otherwise.
In practice we replace $\mathcal{M}_{n+2}$ in Equation~(\ref{eq:primitive}) with
the recursion relation to obtain a more stable expression for $\mathcal{P}(\bvec{G}_n)$:
\begin{proof}{pofgn_v02}\label{eq:PofGn_v02}
\mathcal{P}(\bvec{G}_n) = 2r^2 \mathcal{M}_n - \frac{n}{n+2}\left((1-r^2-b^2)\mathcal{M}_n+(1-(b-r)^2)((b+r)^2-1)\mathcal{M}_{n-2}\right).
\end{proof}

Note that these recursion relations involve every fourth term, so we need to compute
the first four terms analytically.  These are given by:
\begin{proof}{Mn_initial}
\mathcal{M}_0 &= \kappa_0,\nonumber\\
\mathcal{M}_1 &= 2 (4br)^{1/2} \left[E(k^2)-(1-k^2)K(k^2)\right],\nonumber\\
\mathcal{M}_2 &= 4br\left[(k^2-\tfrac{1}{2}) \kappa_0 + k\sqrt{1-k^2}\right],\nonumber\\
\mathcal{M}_3 &= \tfrac{2}{3}(4br)^{3/2} \left[(4k^2-2)E(k^2)+(3k^2-2)(k^2-1)K(k^2)\right],
\end{proof}
for $k^2 \le 1$, while for $k^2 > 1$,
\begin{proof}{Mn_initial}
\mathcal{M}_0 &= \pi,\nonumber\\
\mathcal{M}_1 &= 2(1-(r-b)^2)^{1/2}E(k^{-2}),\nonumber\\
\mathcal{M}_2 &= \pi(1-b^2-r^2),\nonumber\\
\mathcal{M}_3 &= \tfrac{2}{3}(4br)^{3/2} k^3\left[2(2-k^{-2})E(k^{-2})-(1-k^{-2})K(k^{-2})\right].
\end{proof}
We re-express the elliptic integrals for $n=1$ and $n=3$ in terms of the cel integrals which
were already computed for the linear limb darkening case,
\begin{proof}{Mn_cel}
\mathcal{M}_1 &= 2 (4br)^{1/2} k^2 \mathrm{cel}(k_c,1,1,0),\nonumber\\
\mathcal{M}_3 &= \tfrac{2}{3}(4br)^{3/2}k^2 \left[ \mathrm{cel}(k_c,1,1,k_c^2)+(3k^2-2) \mathrm{cel}(k_c,1,1,0)\right],
\end{proof}
for $k^2 \le 1$, while for $k^2 > 1$,
\begin{proof}{Mn_cel}
\mathcal{M}_1 &= 2(1-(r-b)^2)^{1/2} \mathrm{cel}(k_c,1,1,k_c^2),\nonumber\\
\mathcal{M}_3 &= \tfrac{2}{3}(1-(b-r)^2)^{3/2} \left[(3-2k^{-2}) \mathrm{cel}(k_c,1,1,k_c^2)+k^{-2} \mathrm{cel}(k_c,1,1,0\right].
\end{proof}
where, as before, $k_c = \sqrt{1-k^2}$ for $k^2 \le 1$, and $k_c = \sqrt{1-k^{-2}}$ for $k^2 > 1$.

For downward recursion, we compute the top four $\mathcal{M}_n$ expressions, $N-3 \le n \le N$,
in terms of series expansions.
When $k^2 \le 1$, the integrals may be expressed in terms of the following
Hypergeometric functions and infinite series,
\begin{proof}{Mn_series}\label{eq:Mn_series}
\mathcal{M}_n &= (4br)^{n/2} k^{n+1} \pi^{1/2} \frac{\Gamma{(1+\tfrac{n}{2})}}{\Gamma(\tfrac{3}{2}+\tfrac{n}{2})} \,_2F_1(\tfrac{1}{2},\tfrac{1}{2};\tfrac{3}{2}+\tfrac{n}{2};k^2),\nonumber\\
&= (1-(r-b)^2)^{n/2} k \sum_{j=0}^{j_{max}} \alpha_j k^{2j},\nonumber\\
\alpha_0 &= \sqrt{\pi} \frac{\Gamma(1+\tfrac{n}{2})}{\Gamma(\tfrac{3}{2}+\tfrac{n}{2})},\nonumber\\
\alpha_j &= \alpha_{j-1} \frac{(2j-1)^2}{2j(1+n+2j)}.
\end{proof}
Although $j_{max} = \infty$, in practice we set $j_{max} = 100$, and the series is
truncated when a term goes below a tolerance specified by the numerical precision.

We find that upward recursion in $n$ is more stable for $k^2 > \tfrac{1}{2} $,
while downward recursion is more stable for $k^2 < \tfrac{1}{2}$.  Note that
this differs from \citet{starry}, for which downward recursion was also required for $k^2 > 2$.


\subsection{Analytic derivatives}\label{sec:analytic_derivatives}

The derivatives of $\mathcal{P}(\bvec{G}_n)$ may be expressed simply as functions
of $\mathcal{M}_n$:
\begin{eqnarray}
\frac{\partial \mathcal{P}}{\partial r} &=& 2r \left[(n+2)\mathcal{M}_n - n \mathcal{M}_{n-2}\right],\\
\frac{\partial \mathcal{P}}{\partial b} &=& \frac{n}{b} \left[(r^2+b^2)(\mathcal{M}_n - \mathcal{M}_{n-2})+(r^2-b^2)^2\mathcal{M}_{n-2}\right].
\end{eqnarray}
Since the $\mathcal{M}_n$ integrals are computed for the total flux case,
there is little overhead for computing the derivatives.

For small values of $b$ we find that the derivative with respect to impact parameter 
becomes numerically unstable due to the near cancellation between the two
terms, followed by division by $b$.  To avoid this problem
for small $b$, we have derived an alternative expression which avoids division by $b$
which we utilize when $b < b_c$, where $b_c$ is a (small) cutoff value:
\begin{eqnarray}
\frac{\partial \mathcal{P}}{\partial b} = n \left[b\mathcal{M}_n +(2r^3+b^3-3r^2b-b-3) \mathcal{M}_{n-2}-4r^3\mathcal{N}_{n-2}\right],
\end{eqnarray}
where we have defined a new integral, $\mathcal{N}_n$,
\begin{equation}\label{eq:N_of_n}
\mathcal{N}_n(r,b) = (4br)^{n/2} \int_{-\kappa_0/2}^{\kappa_0/2} (k^2-\sin^2\xi)^{\tfrac{n}{2}} \sin^2{\xi} d\xi,
\end{equation}
which obeys the recursion relation
\begin{proof}{Nn_recursion}
\mathcal{N}_n = \frac{1}{n+2} \left[\mathcal{M}_n + n(1-(b+r)^2) \mathcal{N}_{n-2}\right].
\end{proof}
Since this recursion relation involves every other term, we only need the two lowest terms,
which are given by:
\begin{proof}{Nn_cel}
\mathcal{N}_0 &= \tfrac{1}{2}\kappa_0 - k k_c,\nonumber\\
\mathcal{N}_1 &= \tfrac{2}{3}(4br)^{1/2} k^2 \left[-\mathrm{cel}(k_c,1,1,k_c^2) + 2 \mathrm{cel}(k_c,1,1,0)\right],
\end{proof}
for $k^2 \le 1$ and
\begin{proof}{Nn_cel}
\mathcal{N}_0 &= \frac{\pi}{2},\nonumber\\
\mathcal{N}_1 &=  \tfrac{2}{3}(4br)^{1/2} k \left[2\mathrm{cel}(k_c,1,1,k_c^2) - \mathrm{cel}(k_c,1,1,0)\right],
\end{proof}
for $k^2 > 1$.

{\edited For $k^2 < \tfrac{1}{2}$,} we find the upward recursion to be unstable, and so we evaluate the expressions for $\mathcal{N}_{N}$ and $\mathcal{N}_{N-1}$
with a series solution (as we did for $\mathcal{M}_{N-3}$,\, ...,\, $\mathcal{M}_N$):
\begin{proof}{Nn_series}
    \label{eq:Nn_series}
    \mathcal{N}_n &= (4br)^{n/2} k^{n+3} \frac{\pi^{1/2}}{2} \frac{\Gamma{(1+\tfrac{n}{2})}}{\Gamma(\tfrac{5}{2}+\tfrac{n}{2})} \,_2F_1(\tfrac{1}{2},\tfrac{3}{2};\tfrac{5}{2}+\tfrac{n}{2};k^2) \nonumber\\[0.5em]
                  &= (1-(r-b)^2)^{n/2} k^3 \sum_{j=0}^{j_{max}} \gamma_j k^{2j}, \nonumber\\[0.5em]
    \gamma_0 &= \frac{\sqrt{\pi}}{2} \frac{\Gamma(1+\tfrac{n}{2})}{\Gamma(\tfrac{5}{2}+\tfrac{n}{2})},\nonumber\\[0.5em]
    \gamma_j &= \gamma_{j-1} \frac{(4j^2-1)}{2j(3+n+2j)}.
\end{proof}
We then use downward recursion with the relation
\begin{equation}
\mathcal{N}_n = \frac{(n+4)\mathcal{N}_{n+2} - \mathcal{M}_{n+2}}{(n+2)(1-(b+r)^2)}
\end{equation}
to iterate down to $n=3$, while finally computing $n=1$ and $n=2$ exactly.

Evaluating this additional integral adds further computational expense, but in
practice we only need to compute it for $b < b_c = 10^{-3}$ to obtain similar accuracy
to the other expressions.  This is encountered rarely as it only applies when the
occultor is nearly aligned with the source.

With the computation of $\mathfrak{s}_n = -\mathcal{P}(\bvec{G}_n)$, we then
compute the derivatives of the solution vector as
\begin{eqnarray}
\frac{\partial \mathfrak{s}_n}{\partial r} &= & -\frac{\partial \mathcal{P}(\bvec{G}_n)}{\partial r},\\[0.5em]
\frac{\partial \mathfrak{s}_n}{\partial b} &= & -\frac{\partial \mathcal{P}(\bvec{G}_n)}{\partial b},
\end{eqnarray}
for $2 \le n \le N$, while the $n=0$ and $n=1$ terms are handled separately as in
section \S\ref{sec:reparam}.
The derivatives of the normalized flux, $F$, with respect to $r$ and $b$
are then computed as
\begin{eqnarray}\label{eq:derivatives}
\frac{\partial F}{\partial r} &=& I_0 \sum_{n=0}^N \mathfrak{g}_n \frac{\partial \mathfrak{s}_n}{\partial r},\\[0.5em]
\frac{\partial F}{\partial b} &=& I_0 \sum_{n=0}^N \mathfrak{g}_n \frac{\partial \mathfrak{s}_n}{\partial b} \quad.
\end{eqnarray}
Since the normalization constant, $I_0$,
is independent of $\mathfrak{g}_n$ for $n \ge 2$ (Equation~\ref{eq:normalization}), the derivative of $F$
with respect to $\mathfrak{g}_n$ is trivial:
\begin{eqnarray}
    \frac{\partial F}{\partial \mathfrak{g}_n} &=&  I_0 \mathfrak{s}_n
\end{eqnarray}
for $n \ge 2$. For the first two terms, we differentiate
Equation~(\ref{eq:normalization}) to obtain
\begin{eqnarray}
\frac{\partial F}{\partial \mathfrak{g}_0} &=&  I_0 \mathfrak{s}_0 - \pi I_0 F,\\[0.5em]
\frac{\partial F}{\partial \mathfrak{g}_1} &=&  I_0 \mathfrak{s}_1 - \frac{2\pi}{3} I_0 F\quad.
\end{eqnarray}

The derivatives of the light curve with respect to $\mathbf{u}$ are computed by applying
the chain rule to the derivatives of the coefficients, 
$\frac{\partial \mathfrak{g}_j}{\partial u_i} = \mathcal{A}_{ji}$,
\begin{eqnarray}\label{eq:dFdu}
\frac{\partial F}{\partial u_i} =  \sum_{j} \frac{\partial \mathfrak{g}_j}{\partial u_i}\frac{\partial F}{\partial \mathfrak{g}_j}.
\end{eqnarray}

\pagebreak 

\subsection{Summary}
\label{sec:summary}
In the last several sections, we showed that if we express the specific intensity distribution
on the surface of a spherical body as the series
\begin{align}
\frac{I(\upmu)}{I_0} &= 1 - u_1 (1 - \upmu) - u_2 (1 - \upmu)^2 - ... - u_{N}(1 - \upmu)^{N} \quad,
\end{align}
(see Equation~\ref{eq:polynomialld}),
the total flux observed during an occultation is given by the analytic and closed form
expression
\begin{align}
    \label{eq:occint_greens_summary}
    F &= I_0 \, \mathfrak{s}^\mathsf{T} \mathcal{A} \bvec{u} \quad ,
\end{align}
where $I_0$ is a normalizing constant (Equation~\ref{eq:normalization}),
$\mathfrak{s}^\top$ is the solution vector (Equation~\ref{eq:sn}, with special
cases given by Equations~\ref{eq:s0}, \ref{eq:s1}, and \ref{eq:s2}),
a function of only the impact parameter
$b$ and radius $r$ of the occultor, $\mathcal{A}$ is a change of basis
matrix (Equation~\ref{eq:A}), and $\mathbf{u}$ is the vector of limb
darkening coefficients $(u_0 \ u_1 \ u_2 \ ... \ u_N)^\mathsf{T}$.
Note that in general $\mathfrak{s}^\top$ is time-dependent, as it depends
upon the relative positions of the bodies as a function of time, $b(t)$, while
$\mathcal{A}\mathbf{u}$ is time-independent, and thus the matrix
multiplication only needs to be computed once per light curve.

In addition to the flux, $F$, we give the partial derivatives of $F$ with respect
to $r$, $b$, and $\{u_i\}$ (or alternatively $\{\mathfrak{g}_n\}$) in
equations \ref{eq:derivatives}--\ref{eq:dFdu} which are efficient and
accurate to evaluate.

Usually when fitting a light curve the unocculted flux is not equal
to unity, but is some unknown value which needs to be fit for.  
So, the correct procedure is to multiply $F(t)$ by a parameter which 
represents the unocculted flux.
In this case the derivative with respect to the flux constant is trivially 
equal to $F(t)$, and the derivatives with respect to the other parameters 
must be multiplied by the same factor.

With the description of the light curve computation complete, we next discuss
the integration of the light curve over a finite time step.

\section{Time integration} \label{sec:time}

Given that most observations are made over a finite exposure time,
the integration of the light curve over time is necessary to capture the
change in brightness over the timestep
with high fidelity \citep[e.g.,][]{Kipping2010}. When constructing
a light curve, usually one divides the time integral of the flux (the \emph{fluence})
by the integration time
to obtain the time-averaged flux. For optimizing and inferring the
posterior of model parameters, we would like to compute the derivatives
of the time-averaged flux with respect to the model parameters.

The instantaneous flux is a function of $3+N$ parameters in the Green's basis,
$\{r,b,\mathfrak{g}_n\}$, or $2+N$ parameters in the polynomial in basis, and of
these, only one varies with time, $b(t)$.  Thus, we can compute the
time-dependent flux with a model for $b(t)=b(\bvec{x},t)$, where
$b(\bvec{x},t)$ is a model for the impact parameter as a function of time
and model parameters $\bvec{x}$.  The set of model parameters need to be specified
by a function, which, for example, might be a Keplerian orbit of the two bodies
with respect to one another, or a full dynamical model of an $N$-body system.
To compute the derivatives of the light curve with respect to $\bvec{x}$, the
derivatives of the dynamical model must be computed as well, $\partial{b}/\partial{\bvec{x}}$.

The time-averaged flux, $\overline{F}$, and its derivatives, are given by
\begin{eqnarray}\label{eq:avg_flux}
\overline{F} &=& \frac{1}{\Delta t} \int_{t-\tfrac{1}{2}\Delta t}^{t+\tfrac{1}{2}\Delta t} F(t^\prime) dt^\prime,\cr
\frac{\partial \overline{F}}{\partial r} &=& \frac{1}{\Delta t} \int_{t-\tfrac{1}{2}\Delta t}^{t+\tfrac{1}{2}\Delta t} \frac{\partial F(t^\prime)}{\partial r} dt^\prime,\cr
\frac{\partial \overline{F}}{\partial \mathfrak{g}_i}&=& \frac{1}{\Delta t} \int_{t-\tfrac{1}{2}\Delta t}^{t+\tfrac{1}{2}\Delta t} \frac{\partial F(t^\prime)}{\partial \mathfrak{g}_i} dt^\prime,\cr
\frac{\partial \overline{F}}{\partial \bvec{x}} &=& \frac{1}{\Delta t}
\int_{t-\tfrac{1}{2}\Delta t}^{t+\tfrac{1}{2}\Delta t} \frac{\partial F}{\partial b}\frac{\partial b(t^\prime)}{\partial \bvec{x}} dt^\prime,
\end{eqnarray}
where $t$ is taken to be the mid-point of the transit exposure time, and
$\Delta t$ is the exposure time.

As an example, we choose the approximate transit model $b(t) = (b_0^2 + v^2(t-t_0)^2)^{1/2}$,
which ignores acceleration and curvature during a transit, and thus is valid in the
limit of large orbital separation.  We compute the time-averaged flux and derivatives
with respect to $\bvec{x} = \{t_0,v,b_0\}$ for a length of integration time $\Delta t$.
{\edited For exposures which contain a contact point, we break these up into sub-exposures
between the start, end, and contact points within the exposure, and then separately
carry out the integration over each sub-exposure.  This is required due to the fact
that the flux and derivatives are discontinuous at each of the contact points, and
so the time-integration is most efficient when integrating up to, but not over,
a contact point.  For our simplified transit trajectory, $b(t)$, these contact points
can be computed analytically;  for an eccentric orbit, the contact points may require
numerical methods to identify the times of contact before the sub-exposures can be specified.}

\begin{figure}
    \begin{centering}
    \includegraphics[width=\linewidth]{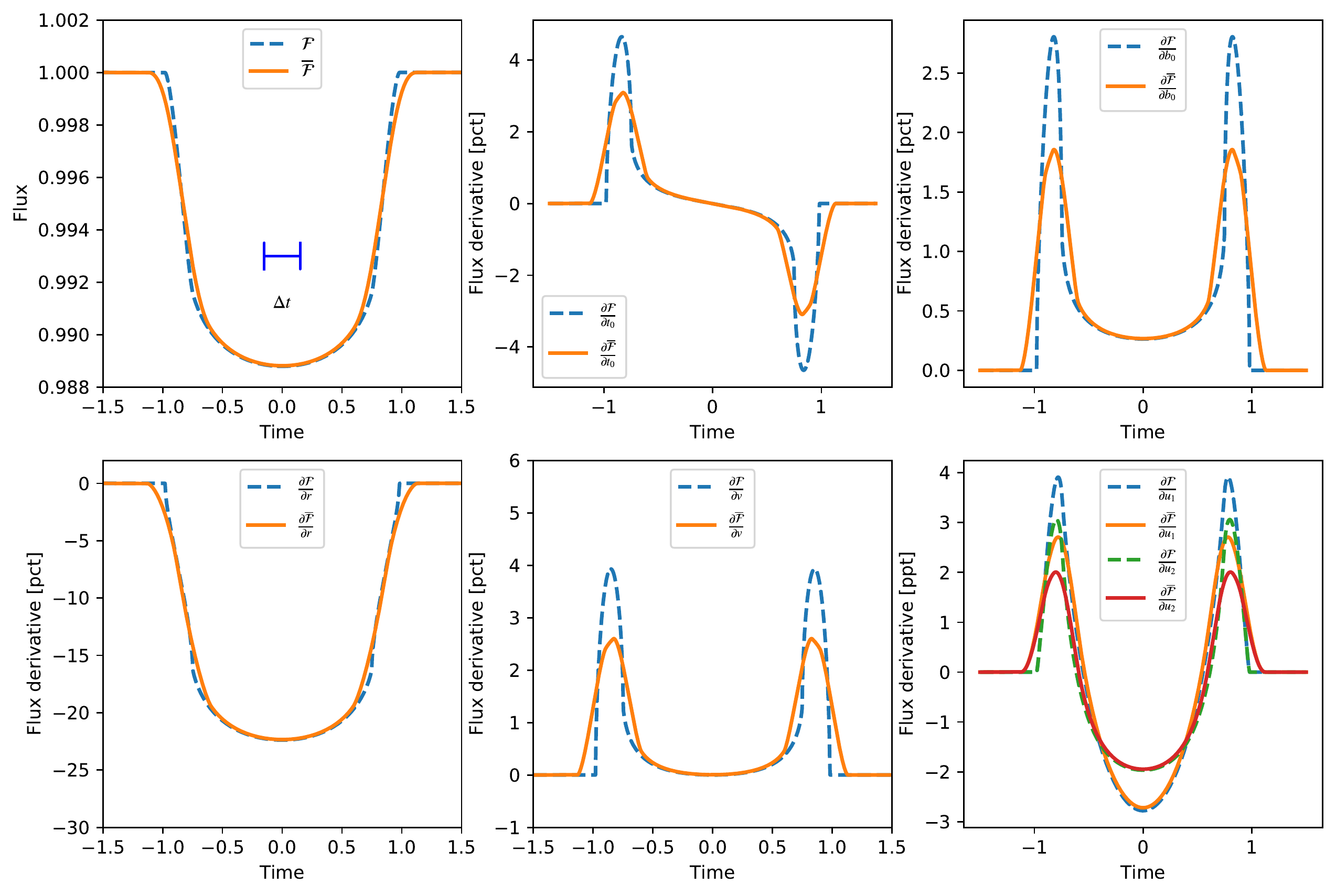}
    \caption{Comparison of the normalized flux and its derivatives with {\edited and
    without time-integration (see solid and dashed lines, respectively)}.  
    The integration time, $\Delta t = 0.3$, is indicated in the
    upper left panel with a horizontal blue line. The derivatives are computed with
    respect to $\{r,t_0,v,b_0,u_1,u_2\}$.  The parameters are given by $r=0.1$,
    $t_0 = 0$, {\edited $v=1$}, $b_0=0.5$, and $u_1=u_2=0.3$. \jlcodelink{integrate_transit_gradient}
    \label{fig:integrated_derivs}}
    \end{centering}
\end{figure}

The integration {\edited of each (sub-)exposure} is carried out with an adaptive Simpson 
quadrature routine \citep{Kuncir1962}.  Since at each point we compute the flux along with
its derivatives, we have developed a vectorized version of this routine
which keeps track of the quadrature separately for each component of the
flux and its partial derivatives, Equations \ref{eq:avg_flux}.  The convergence check for the adaptive
Simpson rule is applied to each component, and when all satisfy the
convergence criterion, the adaptive refinement is terminated.
In practice this algorithm requires specifying a convergence tolerance,
$\epsilon_{tol}$, as well as a maximum number of depths, $D_{max}$, to
allocate memory to store the intermediate results.
Figure \ref{fig:integrated_derivs} shows a
comparison of the derivatives of the time-integrated light curve for this
impact parameter model.

The time-integration smooths both the features of
the light curve, as well as the features of the derivative curves.  The similarity of
the shape of the derivative with respect to $v$ and $b_0$ makes it apparent
that there may be partial degeneracies between the impact parameter and duration
of a transit, which can make the impact parameter more difficult to measure, especially
when the exposure time is longer than the time of ingress/egress.  Likewise, the
derivatives with respect to the two limb darkening parameters have a similar shape,
which explains why in some cases it can be difficult to constrain both parameters.

{\edited In practice the $\epsilon_{tol}$ parameter controls the average number of evaluations
per exposure time, while the tolerance achieved is typically $<10 \epsilon_{tol}$.
Figure \ref{fig:integration_precision} shows the maximum numerical error achieved
for computations with $\epsilon_{tol} = (10^{-4},10^{-6},10^{-8},10^{-10},10^{-12},10^{-14})$
relative to a precision of $\epsilon_{tol} = 10^{-16}$.    The
computed model has $10^4$ exposures for quadratic limb-darkening with the same parameters 
as in Figure \ref{fig:integrated_derivs}
(note that these exposures overlap in time;  in practice many fewer exposures would
be required to compute this light curve).  In computing the time-integrated light
curves, we integrate over the difference of the flux minus one, so that shallow
transit depths will not lose precision.  In Figure \ref{fig:integration_precision},
the achieved precision is plotted versus the average number of evaluations per exposure
for $\overline{F}(t)-1$ and for each of the derivatives.  In all cases but the highest precision,
the flux and all derivatives achieve a precision which is better than $10 \epsilon_{tol}$.
For the highest tolerance case, $\epsilon_{tol} = 10^{-14}$, we find that the
precision exceeds this value slightly;  this is likely due to the model reaching
the limit of double-precision.

Also plotted in Figure \ref{fig:integration_precision} is ten times the
tolerance versus the evaluation time
per exposure relative to the time for a single evaluation per exposure
(dashed line).  This
curve falls to the right of the number ratio line (dotted line) by about 
a factor of $\sim 2$
for high tolerance ($\epsilon_{tol} = 10^{-4}$), to about a factor of $1.3$
for low tolerance ($\epsilon_{tol} = 10^{-14}$).  Thus, the time per
evaluation for the adaptive time-integrated flux and derivatives exceeds the 
expectation given a single evaluation per exposure.  This is likely due to 
the fact that the model computation takes longer for some parameter values 
than others, while the adaptive integration tends to concentrate the 
evaluations at the parameters which are more expensive to evaluate.  In 
addition there may be computation overhead from the adaptive Simpson integration 
function.

\begin{figure}
    \begin{centering}
    \includegraphics[width=\linewidth]{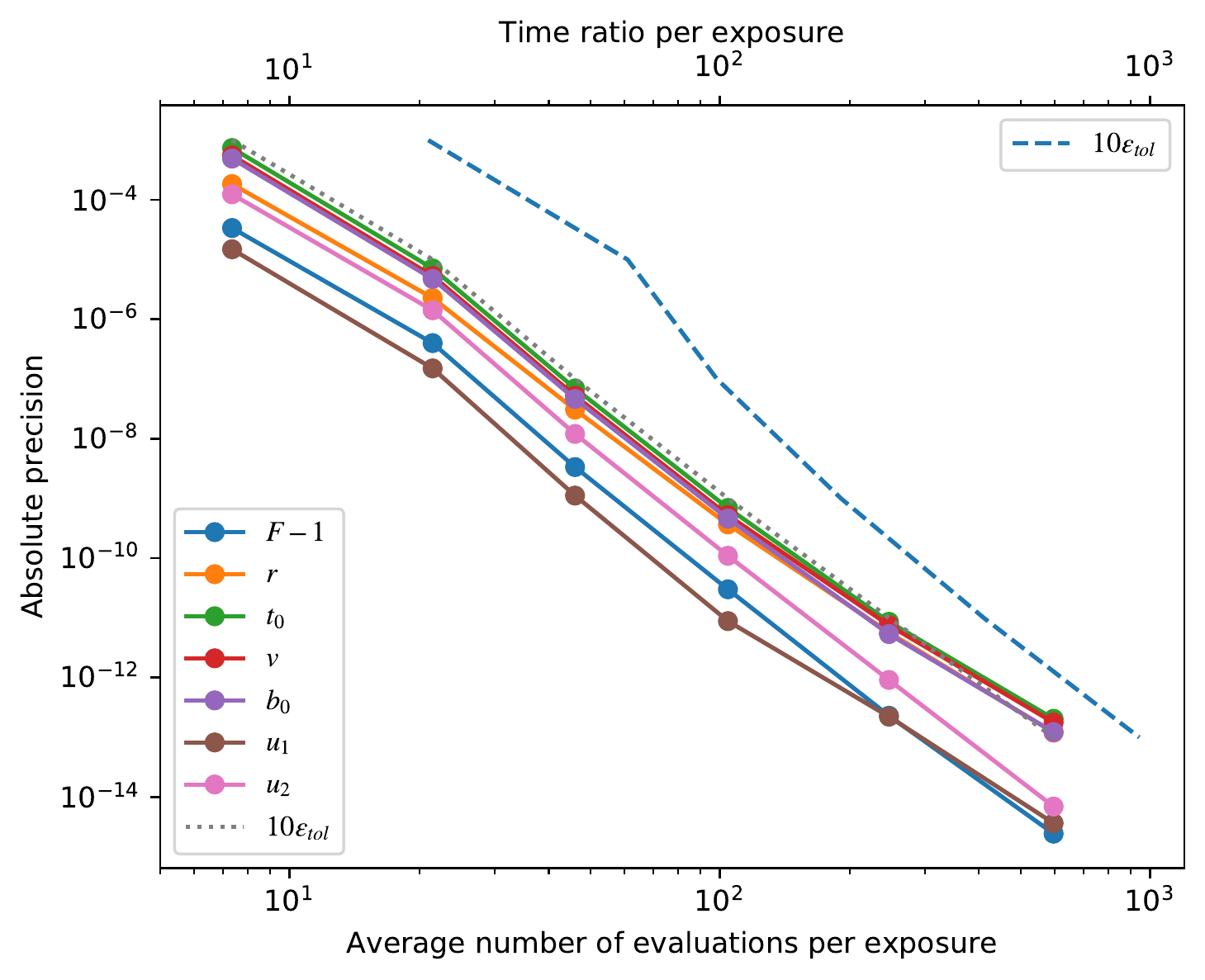}
    \caption{Precision achieved for the time-integrated flux (minus one)
    versus the average number of evaluations per exposure.  The colored lines
    show the precision achieved for $\overline{F}(t)-1$ and for each of the derivatives.
    For comparison, the dotted line shows 10 times the tolerance, $\epsilon_{tol}$.
    The dashed line uses as the abscissa the ratio of the total light curve evaluation 
    time per exposure relative to a single evaluation per exposure. \jlcodelink{profile_integrate_transit}
    \label{fig:integration_precision}}
    \end{centering}
\end{figure}

} 

\pagebreak 

\section{Non-linear limb darkening}\label{sec:nonlinear}

\citet{Claret2000} introduced a ``non-linear'' limb darkening model which
was found to be an effective model for describing the limb darkening functions
which are produced by models of stellar atmospheres.
Although we can only model limb darkening in integer powers of $\upmu$,
we can use a high order polynomial model as an alternative limb darkening model.


We have computed an example non-linear light curve with $r=0.1$ and
$c_1=c_2=c_3=c_4=0.2$, and then fit it with the polynomial limb-darkening
model with increasing orders of the polynomial approximation.  The non-linear light curve model
we computed numerically as the analytic expressions in \citet{MandelAgol2002}
are in terms of hypergeometric functions which are expensive to evaluate.
We numerically compute the non-linear light curve with a ``layer-cake'' model in which sums of
layers of surface brightness with a grid of increasing radii are added together
to approximate the lightcurve;  this
is the approach taken in the numerical model used to compute the
non-linear limb darkening light curves in the code of \citet{MandelAgol2002},
and it is analogous to the approach taken by \citet{Kreidberg2015} for
computing models with arbitrary limb darkening profiles.

\begin{figure}
    \begin{centering}
    \includegraphics[width=0.8\linewidth]{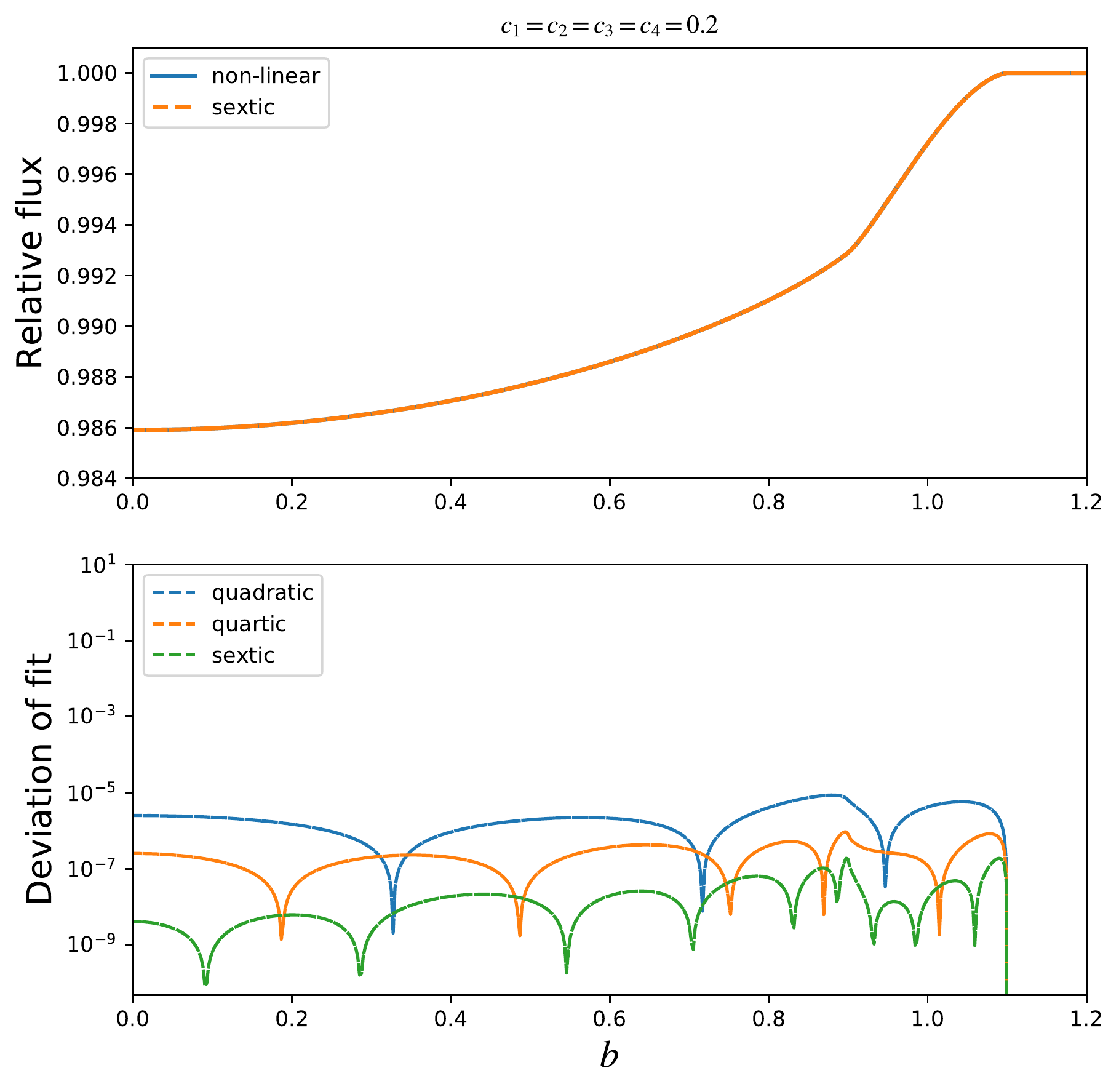}
    \caption{Comparison of the non-linear limb darkening with polynomial
    fits of various orders.
    \jlcodelink{occult_nonlinear}
    \label{fig:nonlinear}}
    \end{centering}
\end{figure}

When fitting the non-linear lightcurve with the polynomial model,
we find that the fit improves steadily up until $N=6$ (a sextic
polynomial), while beyond sextic, the RMS improves imperceptibly.
The RMS of the sextic fit for this example is $<5 \times 10^{-7}$
relative to a depth of transit of about 1.4\% (Figure \ref{fig:nonlinear}).

%
%
%
%

This completes the description of the light curve computation, along
with its derivatives.  We now turn to discussing the implementation of
the computation, followed by comparison with existing codes.

\section{Implementation details}
\label{sec:implementation}

There are several details in our implementation of the foregoing equations
which give further speedup of the computation, which we describe in this
section.

When a light curve is computed, there are some computations which only
need to be carried out once, and then can be reused at each time step in the
light curve computation.
We define a structure to hold these variables
which are reused throughout the light curve; we also pre-allocate variables which
are used throughout the computation to avoid the overhead of memory allocation
and garbage collection.  In addition, due to the greater
computational expense of square-roots and divisions, where possible we try
to only compute a square root or division once, storing these in a variable
within the structure, and then reuse these with cheaper multiplication
throughout the computation when needed.  For instance, for many formulae we
require the inverse of an integer, so an array of integer inverses is computed
once and stored, and then accessed as required rather than recomputed.

Once the number of limb darkening
terms, $N$, is specified, then the series coefficients for $\mathcal{M}_n$
and $\mathcal{N}_n$, $\alpha_j$ and $\gamma_j$, are a simple function of $j$
and $n$, and so we compute these coefficients once, and store them in a vector
for $k^2 \le 1$, separately for $N-3$ to $N$ for $\mathcal{M}_n$, and for
$N-1$ and $N$ for $\mathcal{N}_n$.

In addition, once $N$ is specified, then the transformation matrix for
the Jacobian from $\mathfrak{g}_i$ to $u_j$, $\frac{\partial \mathfrak{g}_i}{\partial u_j}=\mathcal{A}$,
remains the same throughout the light curve
computation, so we compute this matrix only once, and then compute the
flux derivative (Equation \ref{eq:dFdu}) with matrix multiplication.
In fact, since the Jacobian matrix for transforming the derivatives from $\mathfrak{g}_i$
to $u_j$ can be expensive to apply, we can carry out the gradient of the
likelihood function with respect to $\mathfrak{g}_i$, and then apply the Jacobian
transformation from $\mathfrak{g}_i$ to $u_j$ only once to obtain the gradient
of the likelihood with respect to the limb darkening parameterization.
In practice, we are usually only concerned with optimizing a likelihood
or computing gradients of a likelihood for Hamiltonian Markov Chain
Monte Carlo, so the derivatives of the particular points in the
light curve with respect to $u_i$ aren't needed.  This results in a
significant computational savings, especially for large $N$.
Transformation to other parameterizations, such as $q_1$ and $q_2$
defined by \citet{Kipping2013} for quadratic limb-darkening, may also be accomplished after the
fact by applying the Jacobian to compute the gradient of the likelihood
in terms of these transformed parameters.

In computing the elliptic integrals, cel, we found that several terms
which appear in the Bartky formalism are repeated amongst all three elliptic
integrals which appear in the expressions for $\mathfrak{s}_1$.  Consequently, we carry
out a parallel computation of these elliptic integrals such that these
repeated terms are only computed once;  this improves the efficiency of
the elliptic integral computations.  Once these elliptic integrals
are computed for $\mathfrak{s}_1$, the elliptic integrals $\mathrm{cel}(k_c,1,1,k_c^2)=E(m_k)$
and $\mathrm{cel}(k_c,1,1,0)=(E(m_k)-(1-m_k)K(m_k))/m_k $ are
stored in the structure and reused for computing $\mathcal{M}_n$
and $\mathcal{N}_n$.



\section{Benchmarking}
\label{sec:benchmark}

\begin{figure}
    \begin{centering}
    \includegraphics[width=0.7\linewidth]{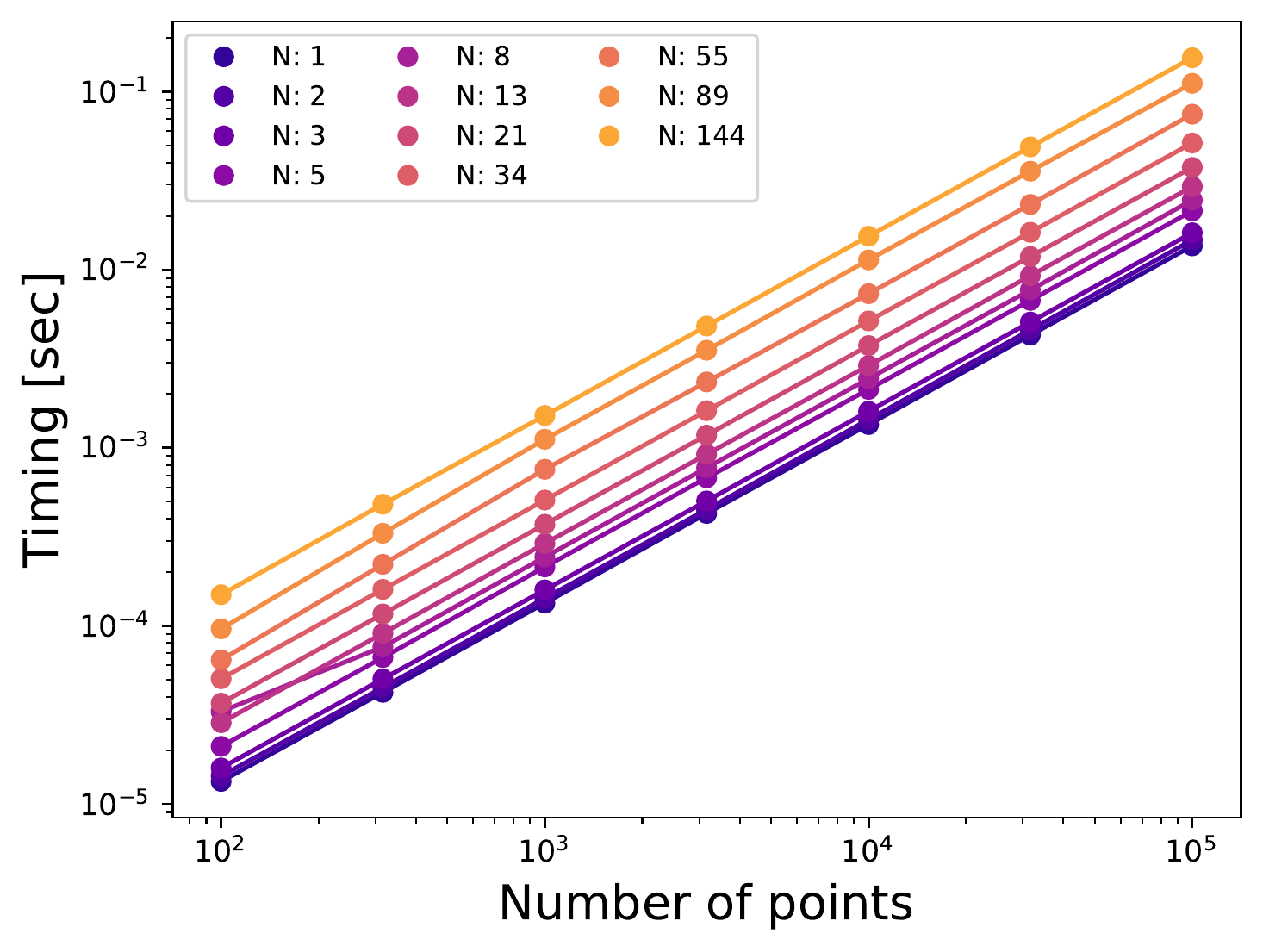}
    \caption{Scaling of the computation time in seconds with the number of
    data points in the light curve for $r=0.1$ with $b$ ranging from $0$ to $1.2$,
    and with the number of limb darkening coefficients, $N$. \jlcodelink{benchmark_transit_poly}
    \label{fig:ncoeff}}
    \end{centering}
\end{figure}

\begin{figure}
    \begin{centering}
    \includegraphics[width=0.7\linewidth]{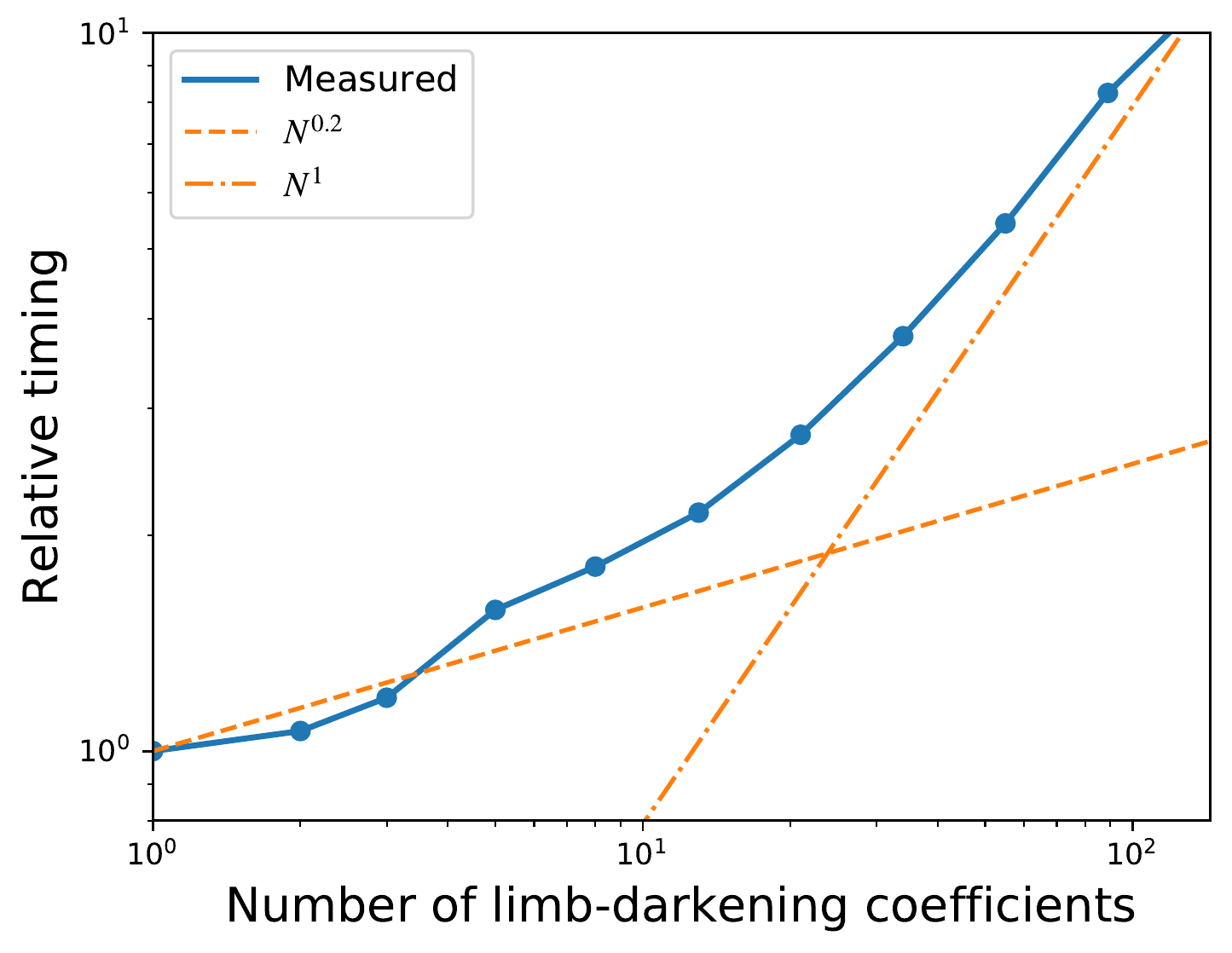}
    \caption{Scaling of the computation time with the number of
    limb darkening coeffients, $N$.  The $y$-axis scales the timing with respect
    to the timing for a single limb darkening coefficient. \jlcodelink{benchmark_transit_poly}
    \label{fig:nlimb}}
    \end{centering}
\end{figure}

We have measured the performance of the limb-darkened light curves
with derivatives as a function of the number of computed data points
and as a function of the number of limb darkening coefficients.  We
have computed the timing for $r=0.1$ and for a number of impact
parameters ranging from $10^2$ to $10^6$, and the number of limb darkening
coefficients ranging from $1$ to $144$.  For each set of timing benchmark
parameters, we carried out nine measurements  of the timing, and
we use the median of these for plotting purposes.  The benchmarking
for the \texttt{Julia} code was carried out with \texttt{v0.7} of
\texttt{Julia} on the \texttt{trusty} Ubuntu environment of Travis-CI%
\footnote{\url{https://travis-ci.org}}
on a 2.3 GHz
2-core machine with 7.5 GB of RAM.
No time-integration/sub-sampling was carried out in this computation.

Figure \ref{fig:ncoeff} shows that the time dependence is linear with the
number of $b$ values (which is equivalent to the number of data points
in the light curve).  The linear scaling with time holds for each value of
the number of limb darkening coefficients.

Figure \ref{fig:nlimb} shows that the time dependence scales approximately
as $N^{0.2-1}$.  As with the number of light-curve points, we have taken
the median over nine measurements for each set of parameters.  We then
scaled the timing to the single-coefficient case, and took a second
median over the number of light curve points.

{\edited
\subsection{Limitation of precision with order of limb-darkening}

We find that the precision of the computation begins to degrade for $N \approx 25-30$.
When computing limb-darkened light curves for very high order limb-darkening,
we have found that the precision can be limited by cancellations which occur
between different orders of the limb darkening.  For standard polynomial
limb-darkening, we find that the $\mathfrak{g}_n$ alternates between very large
positive and negative
values which can end up cancelling to produce a smaller amplitude light curve.
These cancellations can lead to round-off and truncation errors which limit
the precision of the computation.

Figure \ref{fig:g_n_vs_n} shows the values of the $\mathfrak{g}_n$ coefficients
for $u_1 = u_2 = ... = u_{20} = 0.05$ with $N=20$.  Due to the alternating
signs of coefficients in the binomial expansion, the $\mathfrak{g}_n$ values flip
between large negative and large positive values, in this case varying in amplitude
by about seven orders of magnitude from the smallest coefficient to the largest.  
Despite the large values of $\mathfrak{g}_n$, the light curve computed has a
much smaller value with a depth of $\approx 1$\% for a planet with a radius ratio 
of $r=0.1$ due to a fine-tuned cancellation between these large coefficients.
This cancellation is precise for smaller values of $N$ and gradually increases
with $N$.

Figure \ref{fig:fractional_error_vs_order} shows the fractional error found
by computing a light curve at double precision with a lightcurve computed
at \texttt{BigFloat} versus the order of the limb darkening.  Ten trials
were made in which each of the limb-darkening coefficients was randomly
chosen between zero and one, and the sum of their values was normalized
to unity.  In the figure the fractional error is computed relative to
the transit depth for a planet-star radius ratio of $r=0.1$.  For polynomial
limb-darkening order of $N=30$ the fractional error can approach $10^{-6}$.
Consequently we urge caution when using this model with large values of
the polynomial order $N$; the light curve should be checked against a higher 
precision computation for some typical values of the parameters to gauge the 
accuracy of the model.

\begin{figure}
    \begin{centering}
    \includegraphics[width=\linewidth]{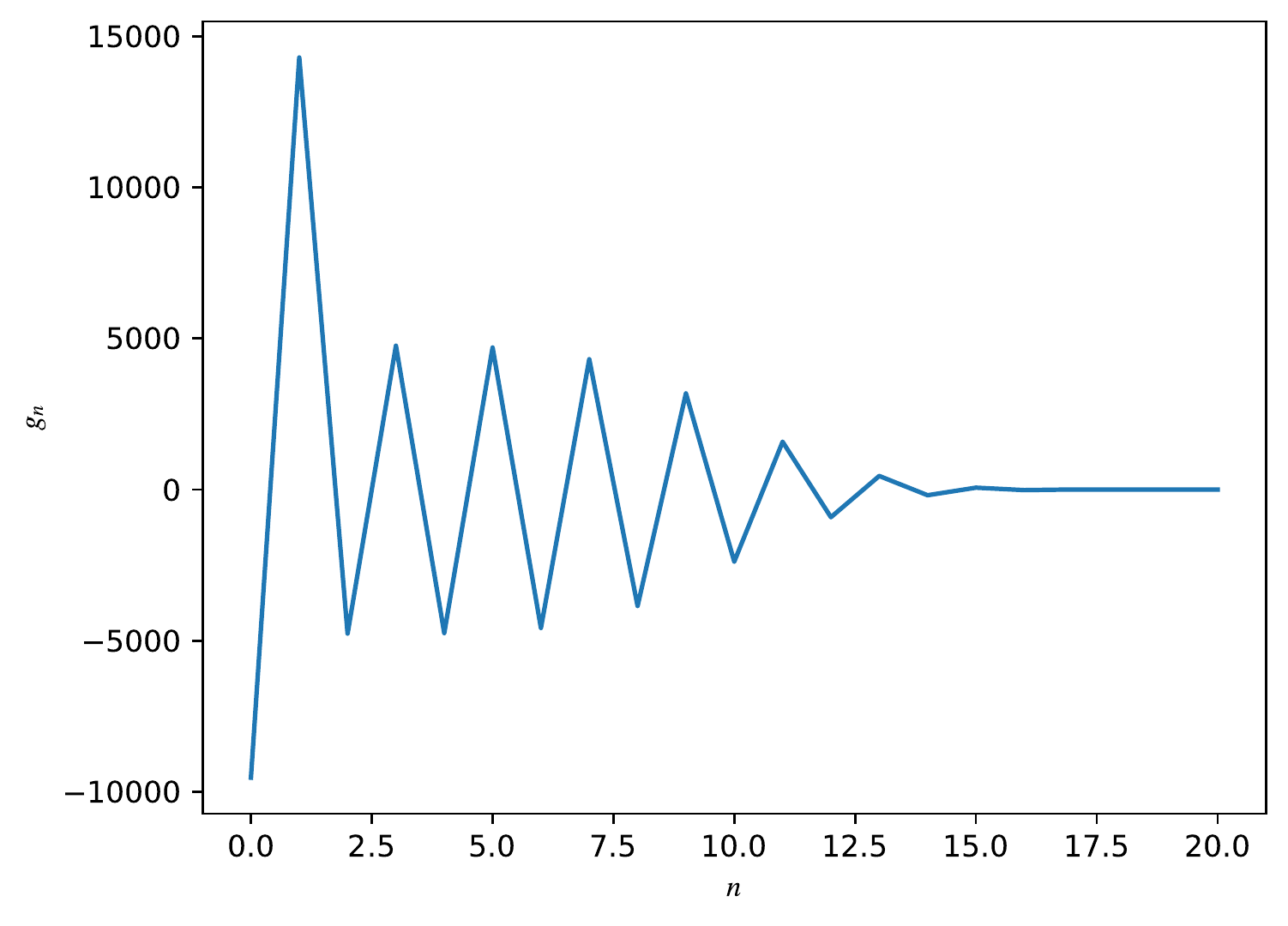}
    \caption{Values of the $\mathfrak{g}_n$ vector versus $n$ for $N=20$
    and for uniform values of $u_n = 1/20$.
    \jlcodelink{plot_g_n}
    \label{fig:g_n_vs_n}}
    \end{centering}
\end{figure}

\begin{figure}
    \begin{centering}
    \includegraphics[width=\linewidth]{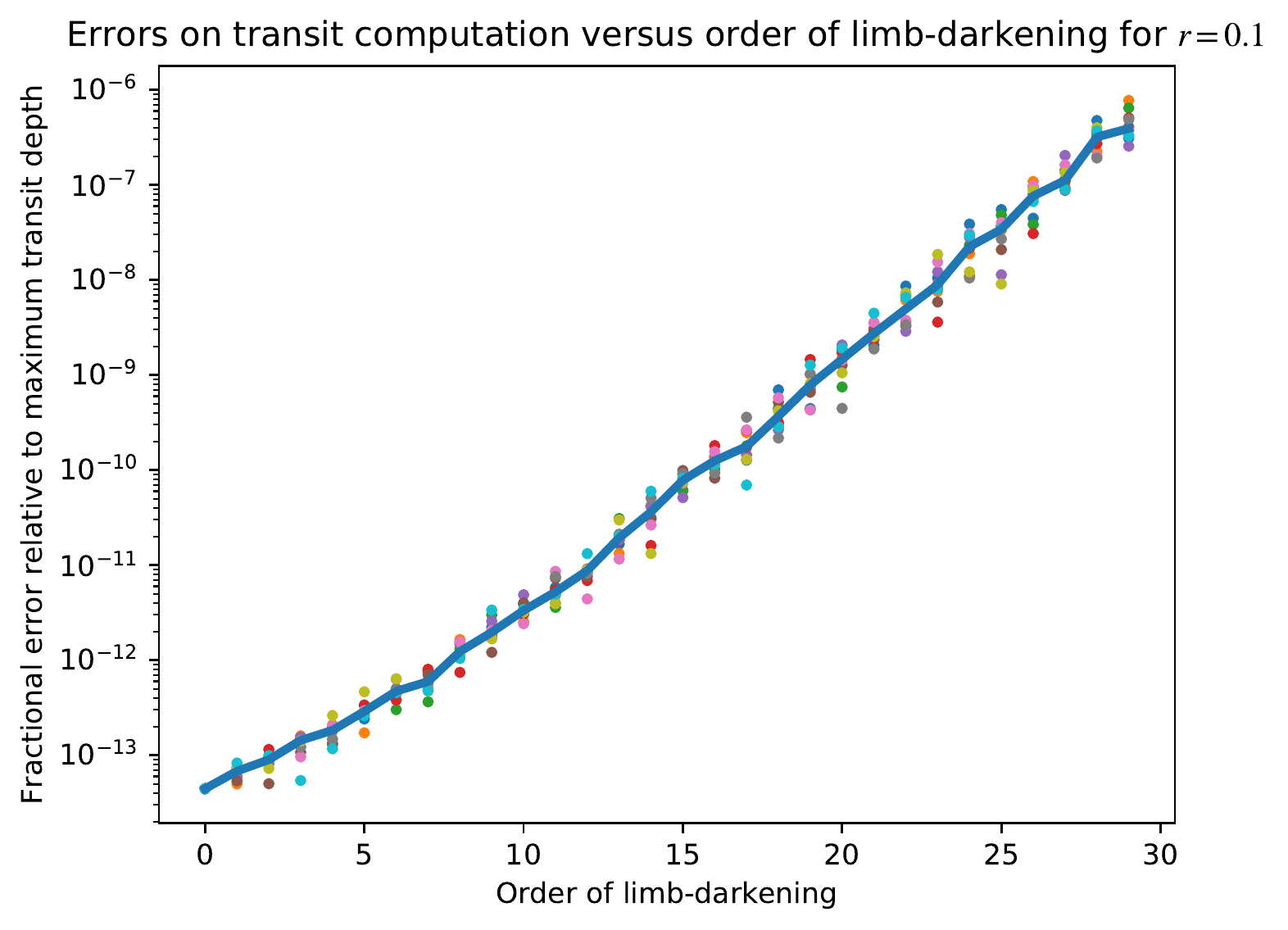}
    \caption{The fractional error on a light curve computed as a function of
    $N$, the order of the limb-darkening.  The dotted points show the results of
    the maximum fractional error (relative to the depth of the transit) for ten
    different random realizations of the $u_n$ limb-darkening coefficients.  The
    thick solid curve shows the median of these ten realizations.
    \jlcodelink{comp_order}
    \label{fig:fractional_error_vs_order}}
    \end{centering}
\end{figure}

}

\section{Comparison with prior work} \label{sec:comparison}

In this section we compare our computations with existing code in terms
of accuracy and speed. We compare both the \texttt{Julia} version
of our code and an implementation of our algorithms
in the \texttt{starry} package, with and without the computation
of gradients. To ensure a fair comparison between the codes, we
perform all calculations on a single core without multi-threading
or interpolation over a pre-computed grid, which is an option
in some codes.

\subsection{Comparison with Mandel \& Agol (2002)}

\begin{figure}[t!]
    \begin{centering}
    \includegraphics[width=0.9\linewidth]{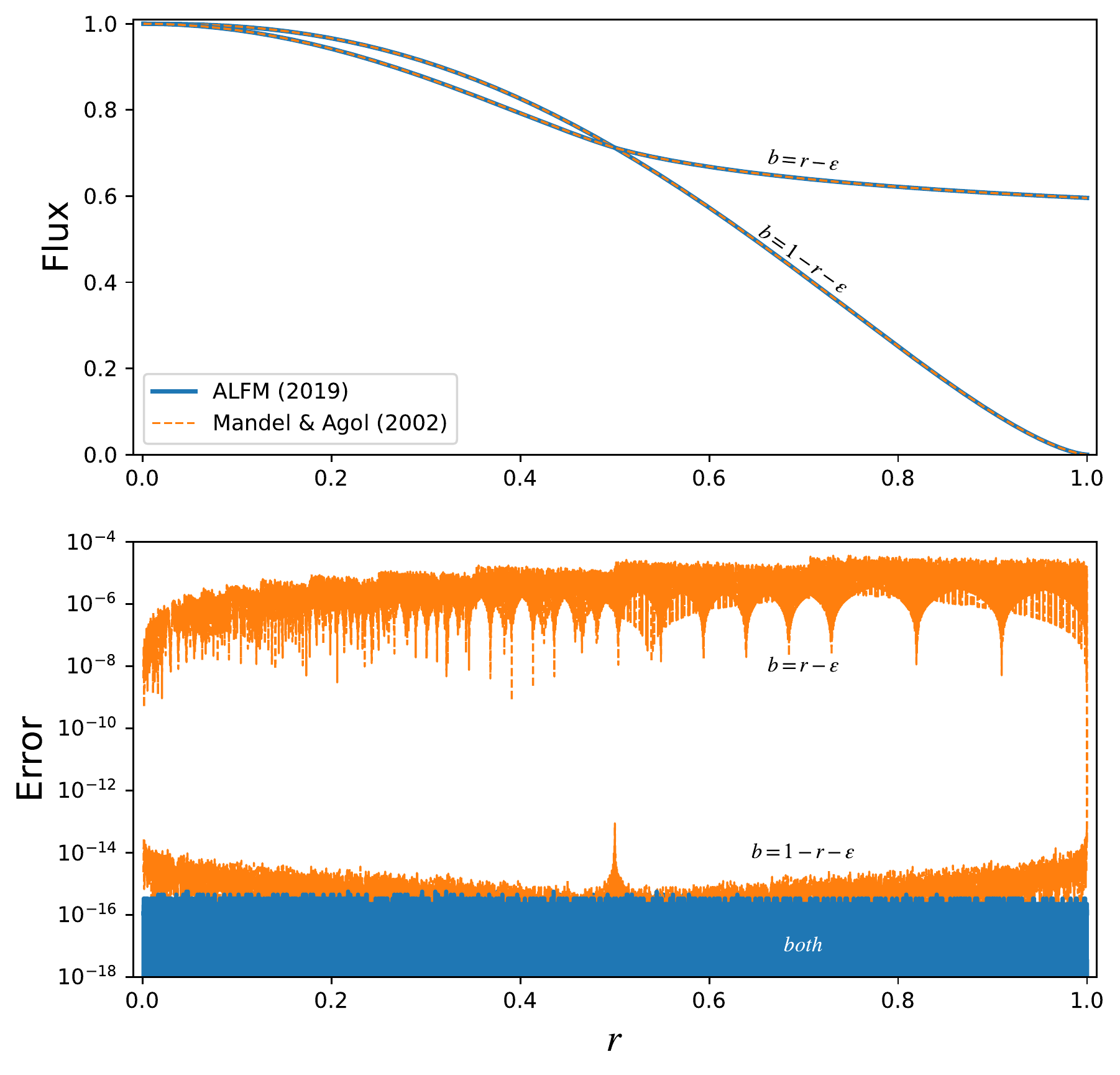}
    \caption{Comparison of Mandel \& Agol (2002) with \thiswork.
    \jlcodelink{compare_MA2002}
    \label{fig:compareMA}}
    \end{centering}
\end{figure}

For uniform, linear or quadratic limb darkening, the \texttt{IDL} package \texttt{EXOFAST} 
improved upon the speed of the widely used computation by \citet{MandelAgol2002} 
by utilizing the \citet{Bulirsch1965a,Bulirsch1965b} expressions for the complete 
elliptic integral of the third kind which is needed for the linear case
\citep{Eastman2013}.  \texttt{EXOFAST} also uses a series
approximation for the complete elliptic integrals of the first and second
kind \citep{Hastings1955}. These three elliptic integrals, especially the third
kind, are the bottleneck in the computation, and the Bulirsch version is faster
than widely used Carlson implementation of elliptic integrals \citep{Carlson1979}.

We have carried out a numerical comparison of the \texttt{EXOFAST} implementation
of the \citet{MandelAgol2002} formulae for the linear case ($u_1=1$), and find 
that the most severe errors occur for $b = r \pm \epsilon$.  Figure \ref{fig:compareMA} 
shows the computed models and the errors as a function of $r$ for $b=1-r-\epsilon$
and $b=r-\epsilon$, with $\epsilon = 10^{-12}$ (the results look very
similar with $+\epsilon$, so we have only plotted one case for clarity).
In the $b\approx 1-r$ case (near second and third contacts), the errors
are larger than our new expression, reaching $\approx 10^{-10}$ for
$r = 1$.  However, the errors become much more severe in the $b \approx r$
case.  For $b=r \pm 10^{-12}$, the errors grow to $10^{-4}$, and continue
to grow as $b$ gets closer to $r$.  No such instability occurs for
our new expressions, demonstrating their utility in all regions of
parameter space.

A speed comparison for a transit computed with $10^7$ data points shows that
the \texttt{Julia} implementation of these routines takes about 55\% of the CPU
time as the \texttt{IDL} \texttt{EXOFAST} implementation without derivatives, and about 65\%
of the computation time when including the derivatives.  Consequently, we
conclude that our new implementation is both faster (by 35-45\%) and more
accurate than the \cite{MandelAgol2002} \texttt{IDL} implementation.  

We note that \texttt{EXOFAST} v2.0 has now been updated to utilize the numerically-stable 
quadratic limb darkening expressions given above, albeit without the computed 
derivatives (Eastman et al., submitted).

\subsection{Derivative comparison with P\'al}

\begin{figure}[p!]
    \begin{centering}
    \includegraphics[width=\linewidth]{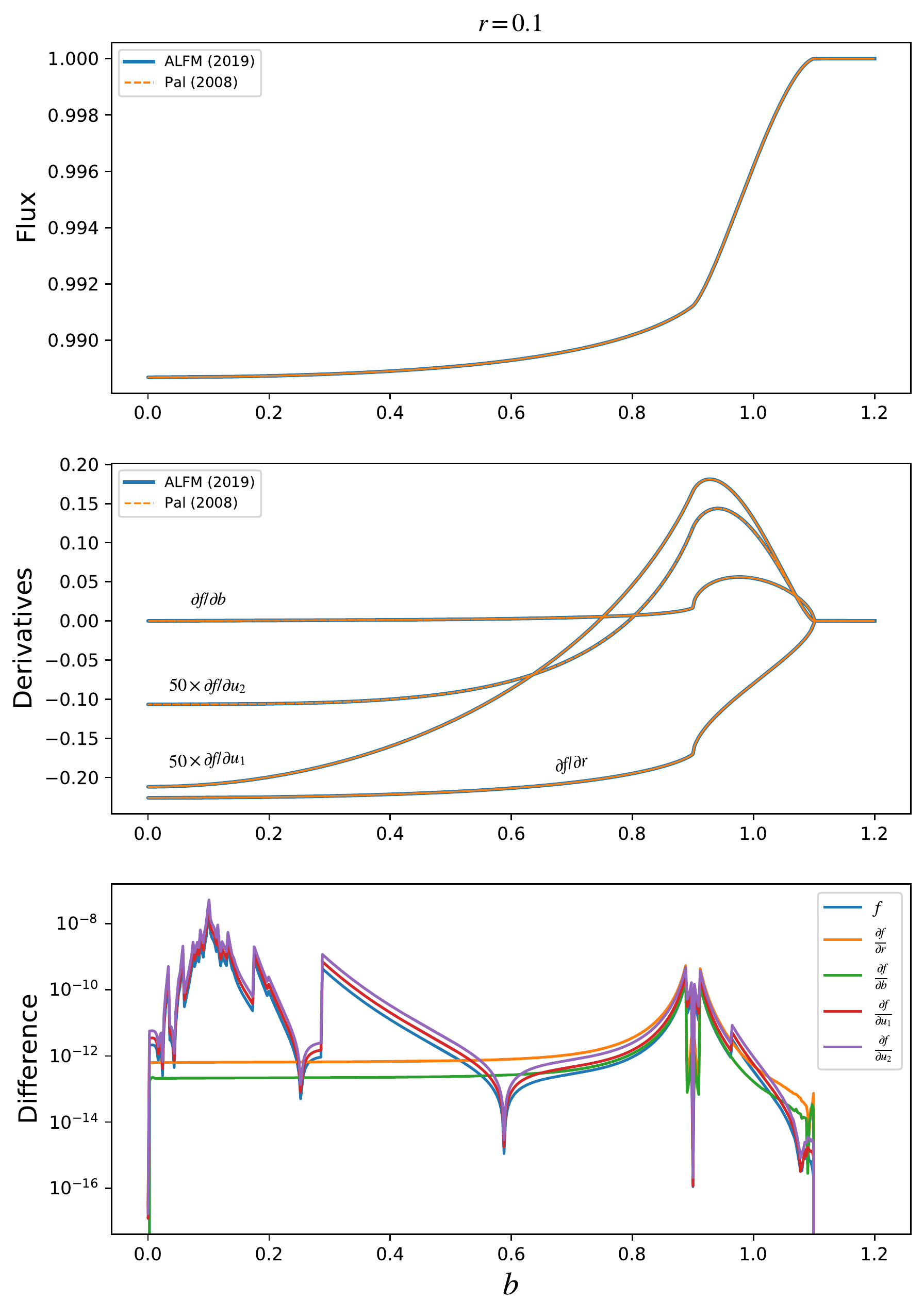}
    \caption{Comparison of \citet{Pal2008} with \thiswork.  The
    coefficients are $u_1=0.2$ and $u_2=0.3$. \jlcodelink{compare_pal}
    \label{fig:Pal_comparison}}
    \end{centering}
\end{figure}

We have computed the quadratic limb-darkened light curve using the \texttt{F77}
code written by Andr\'as P\'al, \texttt{ntiq\_fortran.f}.
Figure \ref{fig:Pal_comparison} shows the results of this comparison.
The light curve models agree quite well, as do the derivatives, which is
a good check on both codes.  However, we find that the P\'al model only
achieves single precision for the computation, with errors reaching as
much as a few $\times 10^{-8}$ for the flux and the derivatives with
respect to the limb darkening parameters.  {\edited One possible origin
for this difference is that} \citet{Pal2008} uses the
Carlson implementation of elliptic integrals \citep{Carlson1979},
which {\edited in this implementation may be} be both less precise and slower to
evaluate than {\edited our new implementation of the}  \citet{Bulirsch1965a} code 
for computing elliptic integrals.

We have also compared the evaluation speed of our code with P\'al's.
We compiled P\'al's code using \texttt{gfortran -O3}, and found that
the computation of quadratic limb-darkened light curves and
derivatives takes an average of 0.52 seconds to compute $10^6$ models,
while the \texttt{transit\_poly\_struct.jl} takes an average of 0.16 seconds,
giving our \texttt{Julia} code a 70\% speed advantage over the Fortran code.

\subsection{Comparison to \texttt{batman}}

\begin{figure}[p!]
    \begin{centering}
    \includegraphics[width=0.95\linewidth]{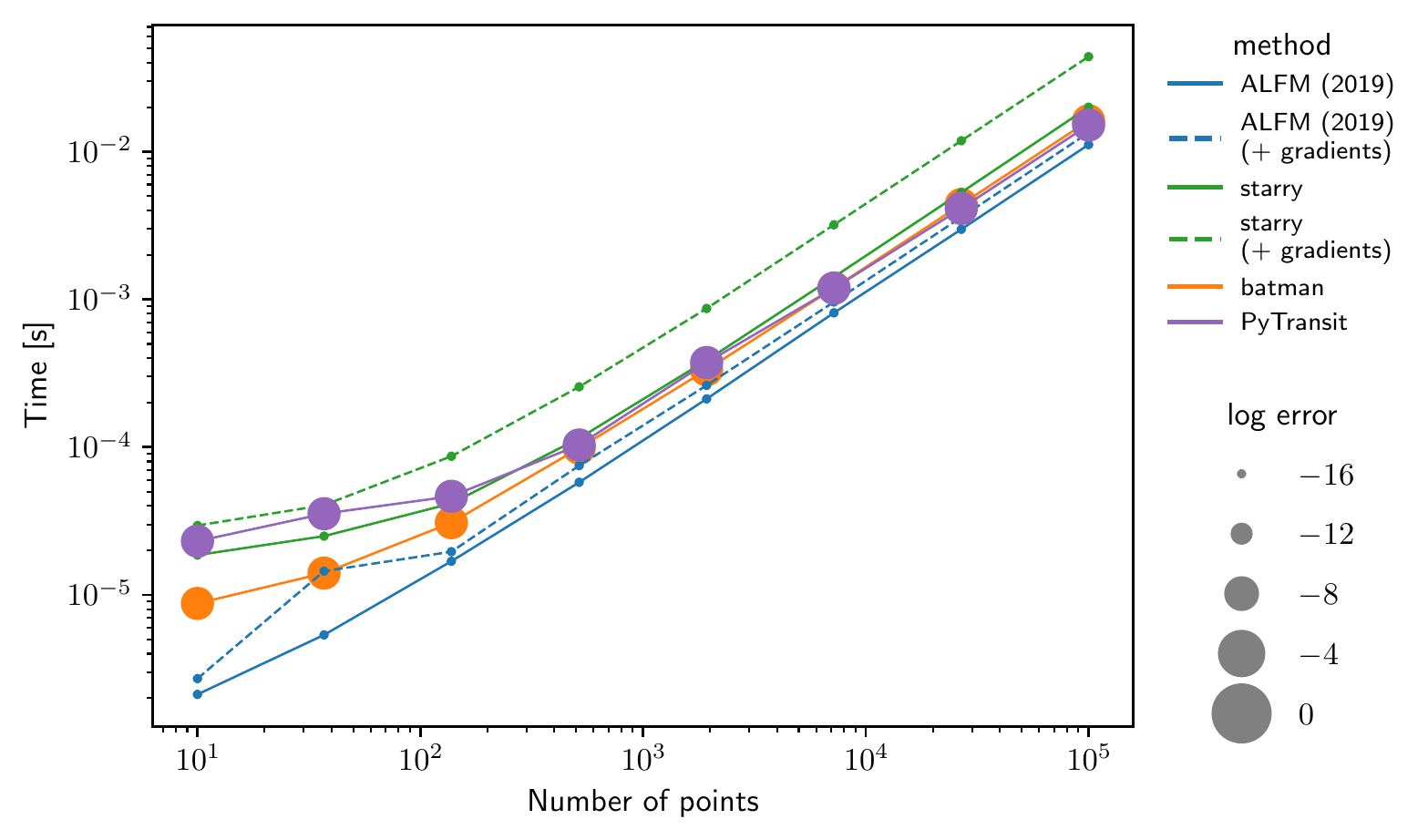}
    \caption{Comparison of \citet{Kreidberg2015} (orange) with \thiswork
             (\texttt{Julia} implementation in blue and \starry implementation
             in green) for a transit across a quadratically limb-darkened star.
             Also shown are points corresponding to the computation using
             \texttt{PyTransit} \citep{Parviainen2015b}. Dashed lines indicate
             computations including the gradients of the flux with respect to
             the radius, impact parameter, and all limb darkening coefficients.
             The $y$-axis corresponds to the evaluation time of the model in
             seconds and the size of the points is proportional to the log of
             the error in the computation relative to a calculation performed
             at 128-bit precision.
             \pycodelink{compare_to_batman}
    \label{fig:batman_comparison}}
    \end{centering}
\end{figure}

\begin{figure}[p!]
    \begin{centering}
    \includegraphics[width=0.95\linewidth]{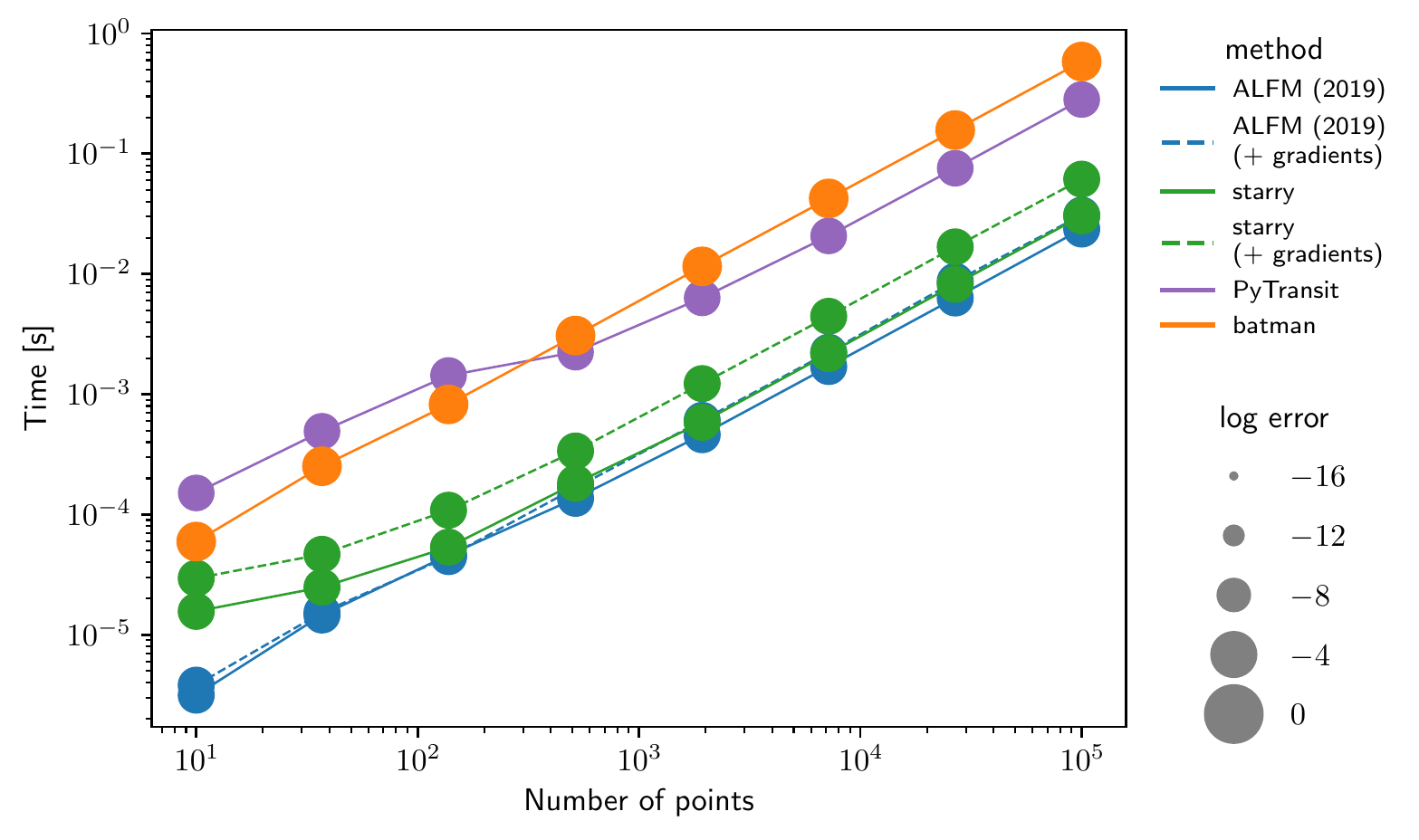}
    \caption{Similar to Figure~\ref{fig:batman_comparison}, but this time comparing
             the computation of a transit across a nonlinearly limb-darkened star.
             \pycodelink{compare_to_batman_nonlinear}
    \label{fig:batman_nonlinear_comparison}}
    \end{centering}
\end{figure}

A \Python implementation of transit light curves which has been widely applied
is the \texttt{batman} package \citep{Kreidberg2015}.  This package
implements a fast C version of the computation, called by \Python for
ease of use.

The \texttt{batman} package computes the quadratic limb darkening model
of \citet{MandelAgol2002}, and uses the same approach for computing
the complete elliptic integrals as \texttt{EXOFAST}.  We have made a comparison
of our implementation of quadratic limb darkening with \texttt{batman},
which is shown in Figure \ref{fig:batman_comparison}.  Without computing
derivatives, our approach (as implemented in \texttt{Julia}; blue)
takes about 60\% of the time of \texttt{batman} (orange);
with derivatives, the two are comparable in speed. The implementation
of our algorithm in \starry (green) is similar in speed to \texttt{batman}
without derivatives, and about a factor of 2 slower than \texttt{batman}
when derivatives are computed. Both the \texttt{Julia} and \starry
implementations have errors close to {\edited double} precision and are therefore
many orders of magnitude more precise than \batman.

Next, we ran a comparison with the non-linear limb darkening model which
proves to be a better fit than the quadratic model to both simulated
and observed stellar atmospheres.  The \texttt{batman} code carries out
a numerical integration over the surface brightness as a function of
radius over the stellar disk, which requires additional computational
time and limits the precision.  We have carried out a fit to the
non-linear limb darkening profile with $c_1=c_2=c_3=c_4=0.2$ with a
polynomial limb darkening model with $N=15$ (\S \ref{sec:nonlinear}).  
We then ran a timing comparison between the \texttt{batman} model and the 
polynomial model, and we find that the polynomial model is about 7 times
more accurate and 25 times faster (\texttt{Julia})
and 20 times faster (\starry) to evaluate compared with \texttt{batman} (Figure
\ref{fig:batman_nonlinear_comparison}).

\subsection{Comparison to \texttt{PyTransit}} \label{sec:comparison_pytransit}

\begin{figure}[t!]
    \begin{centering}
    \includegraphics[width=\linewidth]{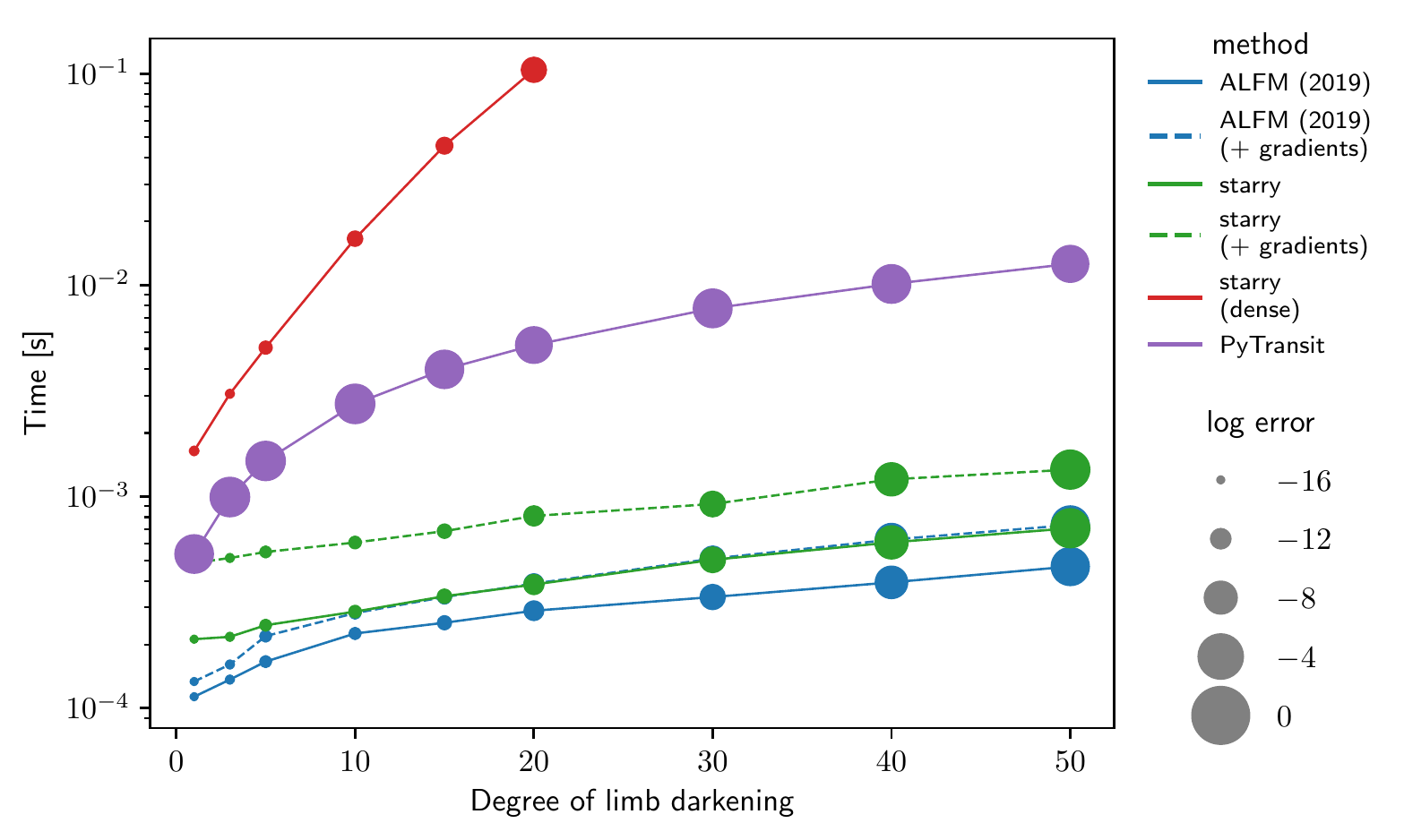}
    \caption{Comparison of the algorithm presented in this work (blue and
             green points) to that of
             \citet{Gimenez2006} for the computation of a transit across a
             star with polynomial limb darkening, as implemented in the
             \texttt{PyTransit} package \citep{Parviainen2015b}
             (purple points), as a function
             of the order of the polynomial. As before, the $y$-axis
             corresponds to the evaluation time and the size of the points
             to the log of the error in the computation. For reference, the
             red points correspond to the evaluation using a naive implementation
             of the full spherical harmonic formalism of \citet{starry}, which
             this paper has improved upon.
             \pycodelink{compare_to_gimenez}
    \label{fig:gimenez_comparison}}
    \end{centering}
\end{figure}

Another popular implementation of transit light curve computation is the
\texttt{PyTransit} code \citep{Parviainen2015b}. We {\edited have} included points corresponding
to this code in Figures~\ref{fig:batman_comparison} and \ref{fig:batman_nonlinear_comparison},
and in general find that it is comparable to \batman in both evaluation time
and accuracy. However, unlike \batman, \texttt{PyTransit} implements the
algorithm of \citet{Gimenez2006} for polynomial limb darkening (Equation~\ref{eq:gimenez}).

In Figure~\ref{fig:gimenez_comparison} we therefore compare our implementation
to {\edited the Gim\'enez algorithm implementation} in \texttt{PyTransit} as a function 
of the degree of limb darkening.
We find our algorithm to be approximately between 5 (for low-order limb darkening)
and 30 (for high-order limb darkening) times faster, and many orders of magnitude
more precise {\edited for low order limb-darkening, while gradually degrading in precision
to higher order limb-darkening to become comparable at very high orders ($N=50$)}. 
{\edited Even when computing derivatives, our \texttt{Julia} code is still faster by
a about a factor of 2.5 for low-order limb-darkening, $N=0$, increasing in
speed relative to the Gim\'enez algorithm by about an order of magnitude at high-orders,
$N=50$}.

For reference, in Figure~\ref{fig:gimenez_comparison},
we also plot the evaluation time when computing the light curve using the
spherical harmonic formalism of \cite{starry}. Because the algorithm presented
in that paper computes surface integrals via recursions in both the
spherical harmonic degree $l$ and the order $m$ (as it was designed to
solve the occultation problem for arbitrary surface features),
it scales super-quadratically with the degree of limb darkening. That
algorithm is therefore orders of magnitude slower
to evaluate in the case of pure limb darkening ($m=0$ modes only). We have
modified the \starry package to compute light curves
using the formalism in this paper in the case of pure limb darkening.

\section{Discussion}
\label{sec:discussion}

We have presented formulae for the transit (or occultation/eclipse) of a
limb-darkened body with a limb darkening profile which is given by a polynomial
in $\upmu$.  These formulae have multiple assumptions built in:  both bodies
are treated as spherical \citep[but see][]{Seager2002,Hui2002}, so that their projected
sufaces are assumed to be circular \citep[but see][]{Barnes2003,Barnes2004,Barnes2009b,
DobbsDixon2012};  limb darkening is treated as azimuthally-symmetric \citep[but see][]{Barnes2009a};
refraction and any relativistic effects are ignored \citep[but see][]{Sidis2010};
and the edges of both bodies are assumed to have a sharp boundary.
All of these assumptions are violated in every transit event to some extent,
but in the majority of cases these assumptions can yield a sufficiently precise model for
a given signal-to-noise ratio.

Given these assumptions, generally one next assumes a particular functional
form for the limb darkening law \citep{Csizmadia2018}.  The parameterization of
the limb darkening model can impact the precision of the computation
of transit light curves.  A common approach is to derive limb darkening
coefficients for a particular limb darkening model from stellar atmosphere models,
and to either fix these at the tabulated values given an observing band and an
estimate of stellar parameters \citep{Claret2011,Howarth2011}, or at least to place
a prior that the limb darkening parameters should nearly match these values.
This approach can have several pitfalls:  the limb darkening model may not be
sufficiently precise,  the stellar atmosphere model may not be accurate, and
the stellar parameters may not be precise.
In computing limb darkening from stellar atmosphere models,
the spherical nature of limb darkening can affect the transit light curve
\citep{Neilson2013,Neilson2017}, and thus the limb darkening coefficients must be fit
with care \citep{Claret2018}.  Even more importantly, full three dimensional
stellar atmosphere models appear to give a more accurate description of
stellar limb darkening by capturing the structure of the atmosphere under
the influence of granulation \citep{Hayek2012,Magic2015}.  However, any
physical model for a stellar atmosphere has limitations in the fidelity at
which it can model actual stellar atmospheres,
and any modeler can only explore a finite set of parameters (effective temperature,
metallicity, surface gravity, and magnetic field strength).
In practice, then, it may be most robust simply to let the limb darkening parameters
be free parameters, to let the limb darkening model be as flexible as possible,
and to let the limb darkening model be fit along with the radius ratio and
orbital parameters \citep{Csizmadia2012,Espinoza2015}.

Even so, this approach still assumes azimuthal symmetry for the star, while
any model for the surface brightness of a star can only be approximate:
to some extent most stars are convective, rotationally-oblate, spotted, oscillating, flaring,
etc.  The model we have presented, then, will only resemble any given star to
a precision which is limited by the lack of uniformity of the actual stellar
surface.  This begs the question of why a numerically precise model is required
for modelling transit light curves.  The answer is computational accuracy
and stability: this more accurate model can be used over all of parameter space,
without returning spurious results, and the high precision enables computation
of derivatives which are beneficial when optimizing model parameters, computing
the Fisher information matrix, or deriving parameter posteriors with MCMC.

Since we are limited in the knowledge of the properties of any given star,
the discrepancies of an azimuthally-symmetric limb-darkened model can be
treated as a source of noise.
The deviation of the star from the model can be absorbed into noise models that
account for outliers, account for correlations in the noise, or actually
try to model the deviations of the star from azimuthal symmetry, such as
induced by star spots \citep[e.g.][]{SanchisOjeda2011}.


One question is what order of the limb darkening model to choose to
fit the data?  Here we suggest several possibile solutions.  The order of the
limb darkening can be varied until the chi-square no longer improves (subject
to a penalty for the greater freedom in the model, such as Bayesian Information
Criterion).  A high-order limb darkening model can be chosen, with the
coefficients regularized to favor small values;  should the data require
a higher-order model, then the coefficients will increase to accommodate
the data.  The parameterization of the limb darkening with terms with
$\mathfrak{g}_n ((n+2)\upmu^n-\upmu^{n-2})$ for $2 \le n \le N$ may be particularly
convenient for this model in that these terms do not contribute to the
total flux of the star.  A third possibility is to fit stellar atmosphere
models with the polynomial limb darkening model until a sufficient precision
is reached given that warranted by the data, and then to place priors
on the limb darkening parameters, informed by the stellar limb darkening
models.  A fourth approach might be to choose a parameterization with
a small number of free parameters, such as the non-linear ``power-2'' law advocated
by \citet{Maxted2018}, and fit this parameterized limb darkening model
with a high-order polynomial for a given set.  Then, only the non-linear
parameters need to be varied, while the polynomial coefficients will
be a simple function of these non-linear parameters.  In this approach
it should be straightforward to linearize the polynomial limb darkening
model fitting, which ought to yield good computational efficiency.
A limitation of our computational approach is that the precision
begins to degrade significantly for $N \approx 25-30$; however, we anticipate that such
a high order will rarely be required.


\section{Applications}
\label{sec:applications}

We envision that this code will be used for fits to higher precision
transit data, such as gathered by the James Webb Space Telescope \citep[JWST;][]{Beichman2014},
which require an improved model of stellar limb darkening.  Here we discuss
some potential avenues for application of this model.

The derivatives of the time-integrated light curves may be used to revisit the
Fisher information analysis as carried out by \citet{Price2014}, as originally
investigated without time-integration by \citet{Carter2008}.  Accounting
for correlated noise in this analysis will give more plausible estimates
for the impact of stellar variability on the determination of transit
transmission spectroscopy and transit-timing variations \citep{ForemanMackey2017}.
This limit will be encountered as more precise measurements are made by gathering
more photons during a transit.  For example, for some targets, one can expect to
obtain $\sim 10^2$ times as many photons with JWST as collected with Kepler.
With such higher precision, as well as the wavelength-dependence afforded
by several JWST observing modes, one can expect that high fidelity transit
models will be required for making precise measurements of transit parameters.

The detection of transit-timing variations with low-amplitude sinusoidal
variations can make use of the fact that small variations in transit time
can be expanded as a Taylor series to linear order in time so that perturbations
in the transit time are the sum of a periodic component and a constant
times the derivative of the limb-darkened light curve \citep{Ofir2018}.
This approach requires derivatives of the light curve with respect to
time, for which the \citet{MandelAgol2002} computation is too
imprecise near the points of contact, $b \approx r$ and $b \approx 1-r$,
within an impact parameter distance of $10^{-4}$, as shown by \citet{Ofir2018},
who interpolated over these regions with polynomials.
However, our new precise formulae, with derivatives, will be useful
for the perturbative approach to the detection of transit timing
variations, avoiding the numerical errors inherent in the \citet{MandelAgol2002}
model over a narrow range of parameter space.

\section{Conclusions}
\label{sec:conclusions}

We have presented an analytic model for the transits, occultations, and
eclipses of limb-darkened bodies with a polynomial dependence of the limb darkening
on the $z$ component of the stellar surface (or, alternatively, the
cosine of the angle from the sub-stellar point, $\upmu$).  The model is more precise
and accurate than prior models that we have compared to, especially near
special limits such as the points of contact and the coincidence of the edge
of the occultor with the center of the source.  The model also compares favorably in
speed of evaluation, about a factor of three faster than the code of
to \citet{Pal2008}, 5-30 faster than that of Gimenez, a factor of 2-25 faster than \texttt{batman}
(depending on the order of the limb darkening), and 35-45\% faster than \texttt{EXOFAST}.

We expect that this code may be used both as a workhorse model for
general fitting of transit models, as well as a tool for more
specialized applications, such as photodynamical modeling of
interacting planets \citep{Carter2012}, triple stars \citep{Carter2011}, and
transiting circumbinary planets \citep{Doyle2011}.

During the preparation of this paper, a related paper appeared on
the mutual eclipse of multiple bodies \citep{Short2018}.  Their
approach is complementary to ours in that they utilize Green's theorem to
carry out a numerical quadrature for mulitiple limb darkening models using
the approach of \citet{Pal2012}.
Their approach does not yet include the computation of derivatives, but
it does allows for a wider range of limb darkening models than polynomial, it
allows for computation of the Rossiter-McLaughlin effect, and it
carries out the computation for multiple overlapping bodies.

The code presented in this paper is open source and is implemented in \edited{three} 
different ways: one of which is a part
of the \starry package, \texttt{http://github.com/rodluger/starry/}, written
in a combination of \texttt{C++} and \Python, 
\edited{another of which is implemented as the default transit model in the 
\Python-based \exoplanet package,
\texttt{http://github.com/dfm/exoplanet/} (Foreman-Mackey et al., in preparation),}
and a new code written in \texttt{Julia},
\texttt{http://github.com/rodluger/Limbdark.jl/}.  We have also implemented
the {\edited quadratic limb-darkened} flux model in \texttt{IDL}, without derivatives; 
this is also available within the GitHub repository. We welcome usage of these
codes, and contributions to further develop and enhance their capabilities.

All figures in this paper were autogenerated on Travis-CI from the latest
version of our repository. Clickable icons (\,\codeicon\,) next to each figure link
to the source code used to produce them, and icons (\,\prooficon\,) next to the main 
equations link to derivations or numerical proofs. We encourage the community
to adopt similar practices to bolster the accessibility, transparency, and
reproducibility of research in the field.

\acknowledgements

We thank Andr\'as P\'al for sharing his Fortran code, \texttt{ntiq-fortran.f}.
We thank Andr\'as P\'al, Kevin Stevenson, Kai Ueltzh\"offer, Mario Damasso,
Matthew Heising, Robert Morehead, and Laura Kreidberg for pointing out
errors or inaccuracies in the \cite{MandelAgol2002} paper and code, which we have
hopefully rectified in this paper.
EA acknowledges NSF grant AST-1615315, NASA grant NNX13AF62G, and from
the NASA Astrobiology Institute's Virtual Planetary Laboratory Lead Team,
funded through the NASA Astrobiology Institute under solicitation NNH12ZDA002C
and Cooperative Agreement Number NNA13AA93A.  This research was partially 
conducted during the Exostar19 program at the Kavli Institute for Theoretical 
Physics at UC Santa Barbara, which was supported in part by the National 
Science Foundation under Grant No.\ NSF PHY-1748958.

\bibliography{limbdark}

\appendix

\section{Errata for Mandel \& Agol (2002)}

Here are a list of errata for \citet{MandelAgol2002}:
\begin{enumerate}
\item In Equation (7), $\lambda_3$ and $\lambda_4$ should have $2k \rightarrow
2p$ in arguments of the elliptic integrals.

\item In Equation (7), $\lambda_5$ should have $- \frac{2}{3}\Theta(p-1/2)$
at the end.

\item For Case 11 in Table 1, $\eta^d$ should be 1/2, not 1, and
$\lambda^d$ should be zero, not 1.  This mistake affects the code,
but it is never encountered for planets that transit main-sequence
stars since $p<1$.  This typo was discussed in \citet{Eastman2013}.

\item The case $z=1-p$ is missing for $z<p$ (as pointed out by
Pal 2008).

\item There is a $\pi$ missing in the denominator of the second term
on the right hand side of Equation (8).
\end{enumerate}

With the exception of 3, none of these errors affected the publicly
available code.

\section{Derivatives of general complete elliptic integral}
\label{app:cel_derivatives}

In this appendix, we give the derivatives of $cel$ with respect to
the input parameters.


\begin{proof}{cel_derivative}
\frac{\partial \mathrm{cel}(k_c,p,a,b)}{\partial k_c} &=& \frac{-k_c}{p-k_c^2}\left[\mathrm{cel}(k_c,k_c^2,a,b)-\mathrm{cel}(k_c,p,a,b)\right],\\
\frac{\partial \mathrm{cel}(k_c,p,a,b)}{\partial p} &=& \frac{\mathrm{cel}(k_c,p,0,\lambda) +(b-ap)\mathrm{cel}(k_c,1,1-p,k_c^2-p)}{2p(1-p)(p-k_c^2)},\\
\lambda &=& k_c^2(b+ap-2bp)+p(3bp-ap^2-2b),\\
\frac{\partial \mathrm{cel}(k_c,p,a,b)}{\partial a } &=& \mathrm{cel}(k_c,p,1,0),\\
\frac{\partial \mathrm{cel}(k_c,p,a,b)}{\partial b} &=& \mathrm{cel}(k_c,p,0,1).
\end{proof}

\section{Listing of symbols and floating point precisions used in the paper and codebase}

Table \ref{tab:symbols} gives a list of the notation used throughout
the main paper.
{\edited Table \ref{tab:precision} lists the IEEE 754 interchange formats
utilized in four versions of this code (\texttt{Limbdark.jl} and \texttt{starry})
written in \texttt{Julia}, \texttt{IDL}, \texttt{Python} and \texttt{C++}.}

\clearpage

\begin{center}
\renewcommand*{\arraystretch}{1.08}
\begin{longtable}{cll}
\caption{Symbols used in this paper} \label{tab:symbols} \\
\toprule
\multicolumn{1}{c}{\textbf{Symbol}} &
\multicolumn{1}{c}{\textbf{Definition}} &
\multicolumn{1}{c}{\textbf{Reference}} \\
\midrule
\endfirsthead
\multicolumn{3}{c}%
{{\bfseries \tablename\ \thetable{} --} continued from previous page} \\
\toprule
\multicolumn{1}{c}{\textbf{Symbol}} &
\multicolumn{1}{c}{\textbf{Definition}} &
\multicolumn{1}{c}{\textbf{Reference}} \\
\midrule
\endhead
\bottomrule
\endfoot
\bottomrule
\endlastfoot
$a_n$           & Gim\'enez coefficients                & \eq{gimenez}\\
$\mathcal{A}$      & Change of basis matrix:
                 $\mathbf{u}$ to Green's
                 polynomials                            & \eq{A} \\
$\mathcal{A}_1$    & Change of basis matrix:
                 $\mathbf{u}$ to
                 polynomials                            & \eq{A1} \\
$\mathcal{A}_2$    & Change of basis matrix:
                polynonials to Green's polynomials      & \eq{gbasis} \\
$A_{lens}$      & Lens-shaped area of overlap of
                  two circles                           & \eq{MAuniform}\\
$A_{kite}$      & Kite-shaped area b/w center
                  of circles and points of contact		& \eq{Kite_area}\\
$b$             & Impact parameter in units of occulted
                 body's radius                          &  \\
$b_0$             & Minimum impact parameter in time-integrated model & \S\ref{sec:time}\\
$b_c$           & Cutoff for using alternative
                  expression for $d\mathcal{P}/db$      & \S\ref{sec:analytic_derivatives}\\
$c_1-c_4$       & Non-linear limb darkening
                  coefficients                          &  \S\ref{sec:nonlinear}\\
$\bvec{D}\,\wedge$
                & Exterior derivative                   & \eq{DGg} \\
$\mathrm{cel}(k_c,p,a,b)$
                & General complete elliptic
                 integral \citep{Bulirsch1969}          & \eq{cel}\\
$E(\bigdot)$    & Complete elliptic integral of the
                 second kind                            & \eq{elliptic} \\
$F$             & Normalized flux seen by observer      & \eq{occint} \\
$\overline{F}$  & Time-averaged normalized flux         & \eq{avg_flux} \\
$_2F_1$         & Generalized Hypergeometric function   & \eq{Mn_series} \\
$\gbasis$       & Green's basis                         & \eq{greensbasis} \\
$\mathfrak{g}$  & Vector in the basis $\gbasis$         & \\
$\bvec{G}_n$    & Anti-exterior derivative of the
                 $n^\mathrm{th}$
                 term in the Green's basis              & \eq{greens_n} \\
$i$             & Dummy index                           & \\
$I$             & Specific intensity, $I(\x, \y)$       & \\
$I_0$           & Intensity normalization constant      & \eq{normalization} \\
$j$             & Dummy index                           & \\
$k$             & Elliptic parameter                    & \eq{k2} \\
                & Dummy index                           & \\
$k_c$           & $\sqrt{1 - k^2}$                      & \eq{cel} \\
$K(\bigdot)$    & Complete Elliptic integral of the
                  first kind                            & \eq{elliptic} \\
$m_k$           & Elliptic integral parameter           & \S\ref{sec:reparam}\\
$n$             & Order of limb darkening/Green's basis	& \\
$\mathcal{M}_n(r,b)$
                & Integral computed recursively         & \eq{M_of_n}\\
$N$             & Highest order of limb darkening polynomial & \\
$\mathcal{N}_n(r,b)$
                & Integral computed recursively         & \eq{N_of_n}\\
$p$             & Cofficient of $cel$			        & \eq{cel}\\
$\pbasis$       & Polynomial basis                      & \eq{polybasis} \\
$\mathfrak{p}$  & Vector in the basis $\pbasis$         & \\
$q$             & Term in cel identities                & \eq{cel_identities}\\
                & Term in cel $\Lambda$			& \eq{biglam_stable}\\
$\mathcal{P}$   & Primitive integral along perimiter
                 of occultor                            & \eq{primitiveP} \\
$\mathcal{Q}$   & Primitive integral along perimiter
                 of occulted body                       & \eq{primitiveQ} \\
$r$             & Occultor radius in units of occulted
                 body's radius                          & \S\ref{sec:intro} \\
$\bvec{r}$      & Vector for integration over
                 boundary of visible disk               & \eq{greens} \\
$\mathfrak{s}$  & Occultation light curve solution
                 vector                                 & \eq{greens} \\
$t$             & Time variable                         & \S\ref{sec:time}\\
$t_0$           & Central time of transit               & \S\ref{sec:time}\\
$u_1, u_2$      & Quadratic limb darkening coefficients & \eq{quadraticld} \\
$\ubasis$       & Limb darkening basis                  & \eq{ldbasis} \\
$\bvec{u}$      & Vector of limb darkening coefficients
                 in the basis $\ubasis$                 & \S\ref{sec:poly_limbdark} \\
$v$             & Velocity in time-integrated model & \S\ref{sec:time}\\
$\bvec{x}$      & Parameters used in time integration   & \S\ref{sec:time}\\
$\x$            & Cartesian coordinate                  & \eq{xyz} \\
$\y$            & Cartesian coordinate                  & \eq{xyz} \\
$\z$            & Cartesian coordinate,
                 $z = \sqrt{1 - \x^2 - \y^2}$           & \eq{xyz} \\
$\alpha_j$      & Coefficient in series for
                  $\mathcal{M}_n$                       & \eq{Mn_series} \\
$\gamma_j$      & Coefficient in series for
                  $\mathcal{N}_n$                       & \eq{Nn_series} \\
$\Gamma$        & Gamma function                        & \\
$\eta$          & Parameter in quadratic limb
                  darkening term                        & \eq{eta}\\
$\theta$        & Polar angle on star with
                  respect to observer                   & \\
$\Theta$        & Heaviside step function               & \eq{biglam} \\
$\kappa_0$      & Angular position of occultor/occulted
                  intersection point                    & \eq{cosine_formulation} \\
$\kappa_1$      & Angular position of occultor/occulted
                  intersection point                    & \eq{cosine_formulation} \\
$\lambda$       & Angular position of occultor/occulted
                  intersection point                    & \eq{primitiveQdef} \\
                & Term in cel identities                & \eq{cel_identities} \\
$\Lambda^e$       & Term in uniform transit expression       & \eq{MAuniform} \\
$\Lambda$       & \citet{MandelAgol2002} function       & \eq{biglam} \\
$\upmu$         & Cosine of polar angle on star,
                  $\upmu = z$                           & \eq{quadraticld} \\
$\Pi(\bigdot,\bigdot)$
                & Complete elliptic integral of the
                 third kind                             & \eq{elliptic} \\
$\phi$          & Angular position of occultor/occulted
                  intersection point                    & \eq{primitivePdef} \\
$\varphi$      & Dummy integration variable             & \\
$\xi$          & Transformed integration variable       & \eq{greens_transformed}\\
\end{longtable}
\end{center}

\clearpage

{\edited 
\begin{table}
\caption{Floating point standards used in implementations of code. \label{tab:precision}}
\begin{tabular}{llll}
\tableline
Language & Variable type & IEEE 754 interchange format & precision for 1.0\\
\tableline
Julia & Float64 & binary64 & $2^{-52} = 2.22\times 10^{-16}$\\
Julia & BigFloat & binary256 & $2^{-255} = 1.73\times 10^{-77}$ \\
IDL & double & binary64 & $2^{-52} = 2.22 \times 10^{-16}$\\
Python & double & binary64 & $2^{-52} = 2.22 \times 10^{-16}$\\
C++ & double & binary64 & $2^{-52} = 2.22 \times 10^{-16}$\\
\tableline
\end{tabular}
\end{table}}
\end{document}